\documentclass[a4paper,11pt]{article}
\usepackage{jheppub}
\usepackage[T1]{fontenc}
\usepackage{booktabs,array,tabularx,microtype,amsthm,mathtools,float}
\DeclareRobustCommand{\family}[1]{\ensuremath{F_{#1}}}
\newcommand{\cD}{\mathcal D}
\newcommand{\perpgrad}{{\nabla_{\!\perp}}}
\newcommand{\eps}{\varepsilon}
\newcommand{\OO}{\mathcal O}
\newcommand{\Lh}{\mathcal L}
\newcolumntype{Y}{>{\raggedright\arraybackslash}X}
\renewcommand{\afterTocRuleSpace}{}
\allowdisplaybreaks[1]
\theoremstyle{plain}
\newtheorem{proposition}{Proposition}[section]

\renewcommand{\Re}{\operatorname{Re}}
\renewcommand{\Im}{\operatorname{Im}}
\makeatletter
\gdef\@fpheader{}
\makeatother
\title{\rmfamily\fontsize{20.74}{25}\selectfont\bfseries
Stable and causal frames for BDNK hydrodynamics}
\author[1,2,3,4]{Jay Armas}
\affiliation[1]{Institute for Theoretical Physics, University of Amsterdam,
1090 GL Amsterdam, The Netherlands}
\affiliation[2]{Dutch Institute for Emergent Phenomena,
1090 GL Amsterdam, The Netherlands}
\affiliation[3]{Institute for Advanced Study, University of Amsterdam,
Oude Turfmarkt 147, 1012 GC Amsterdam, The Netherlands}
\affiliation[4]{Niels Bohr International Academy, The Niels Bohr Institute,
University of Copenhagen, Blegdamsvej 17, DK-2100 Copenhagen \O, Denmark}
\emailAdd{j.armas@uva.nl}
\abstract{We establish the existence of families of linearly stable and causal Bemfica-Disconzi-Noronha-Kovtun
(BDNK) frames for arbitrary equation of state and set of physical transport coefficients. By developing a method 
based on simplification of Routh-Hurwitz criteria and telegrapher factorisation of polynomials, we provide six
concrete examples of such families, three of which admit conformal limits, for both neutral and charged fluids. These frames have adjustable characteristic speeds and can serve 
as a starting point for numerical implementations of relativistic hydrodynamics in various settings.
In the process we uncover a map between stable and causal frames in neutral fluids and stable and causal frames 
in charged fluids. We also develop an \href{https://jacomearmas.org/research/BDNK}{AI research protocol} that can be used to search for such frames in any hydrodynamic theory.
}
\begin{document}
\maketitle
\setcounter{page}{1}

\section{Overview}\label{sec:overview}

Relativistic hydrodynamics describes the long-wavelength evolution of
conserved energy, momentum and charge. Its applications in nuclear physics
and astrophysics require a consistent treatment of dissipation. However,
constitutive relations truncated at first order in gradients need not
define stable and causal evolution equations when treated as an exact
system~\cite{HiscockLindblom}. This matters in numerical simulations, which also probe
wavelengths outside the hydrodynamic regime: an instability there can
prevent the extraction of long-wavelength observables.

M\"uller-Israel-Stewart (MIS) theories address this problem by promoting
dissipative stresses and currents to additional variables with relaxation
equations~\cite{Muller,IsraelStewart}. Another approach is the density frame
formulation~\cite{ArmasJain}, in which conserved densities define the
hydrodynamic variables on a chosen time foliation. Its linearized theory
is stable, and boost symmetry is recovered to the retained order in the
derivative expansion. This diffusive formulation provides an alternative
organization of relativistic hydrodynamics, without requiring the
truncated equations to be an exactly causal hyperbolic system.

The Bemfica-Disconzi-Noronha-Kovtun (BDNK) formulation instead uses
hydrodynamic frame freedom to obtain causal and stable evolution while
retaining covariant constitutive relations in temperature, chemical
potential and velocity~\cite{BDN2018,Kovtun2019,BDN2019,HoultKovtun,BDN2022}.
Time derivatives in the first-order constitutive relations produce
second-order conservation equations and additional nonhydrodynamic modes.
Suitable frame choices give causal, linearly stable systems, with local well-posedness established
for several constitutive classes.

Numerical studies provide encouraging tests of this formulation. Stable,
convergent evolutions of smooth plane-symmetric conformal flows have been
compared with MIS~\cite{PandyaPretorius}, and conservative finite-volume
methods have been developed for multidimensional flows~\cite{PandyaMostPretorius}.
Nonconformal ideal-gas calculations include heat flow and shock
profiles~\cite{PandyaIdealGas}, while holographically initialized
evolutions test hydrodynamics against microscopic
relaxation~\cite{BantilanBeaFigueras}. Spherically symmetric neutron-star
calculations in the Cowling approximation also exhibit stable evolution
within a restricted parameter range~\cite{ShumEtAl}. These developments motivate further tests of BDNK as an effective
classical description.

The main obstacle to BDNK hydrodynamics is finding linearly stable and causal
frames for a given equation of state and \emph{physical} transport coefficients.\footnote{By \emph{physical} transport coefficients we mean the coefficients that appear in entropy production and Kubo formulae. In the cases studied here these are bulk and shear viscosities, as well as the conductivity.} General conditions for
causality and stability are available, including charged constructions
\cite{HoultKovtun,BDN2022}, necessary and sufficient conditions for generic
charged frames, including non-thermodynamic choices~\cite{AbboudSperanzaNoronha},\footnote{Non-thermodynamic frames were also considered recently in \cite{Armas:2025uyv, Armas:2026mul} in the context of ultrarelativistic limits.} and a strongly hyperbolic class
with a first-order corrected charge current, constraints on the equation of state, and large frame parameters~\cite{SchianchiAbalos}.
General criteria characterize admissible coefficients, but using them still
requires solving complicated inequalities for a given equation of state and set of
physical transport coefficients. 

In this work we provide a concrete strategy for finding such frames for any equation of state (EoS) and 
physical transport coefficients. We take the point of view that, within BDNK theory, the EoS and physical transport coefficients are
the properties that characterise the fluid microscopically and can be obtained from e.g. kinetic theory or holography.
The remaining BDNK coefficients can be freely
chosen~\cite{HoultKovtunMicroscopic}. With this in mind, we brute-force
the reduction of the Routh-Hurwitz stability criteria in order to find
analytically tractable solutions and brute-force the factorisation of
the longitudinal (sound) polynomial to simplify the analysis and find
concrete families of frames that are general enough to accommodate
all EoS and physical transport coefficients.\footnote{\label{fn:existing-frames-overview}The constitutive class of Ref.~\cite{BDN2022} also admits stable and causal choices for every admissible EoS and prescribed physical transport, for neutral fluids and, at nonzero density, for charged fluids as we explicitly show in Appendix~\ref{app:existing-frames}. However, its ideal charge current imposes an additional obstruction: finite coefficients at $n=0$ require vanishing conductivity as already noted in Ref.~\cite{BDN2022}. Our families therefore provide further explicit prescriptions, including non-thermodynamic choices and charged frames that remain regular at zero density.}

In Section~\ref{sec:constitutive} we provide the constitutive relations and basic thermodynamic properties of the fluids considered here. In Section~\ref{sec:construction-strategy}
we explain the construction method, allowing both thermodynamic and non-thermodynamic frames.  In sections~\ref{sec:factorization} and
\ref{sec:catalogue} we provide six families of frames and their properties for both neutral and charged fluids while in Section~\ref{sec:discussion} we conclude with a summary and open questions. In
Appendix~\ref{app:proofs} we give various derivations, checks, and proofs.
Appendix~\ref{app:existing-frames} compares these constructions with existing neutral and ideal-current charged frames.

\subsection{Constitutive relations and physical data}\label{sec:constitutive}
To begin with, we consider a parity-even fluid with one conserved $U(1)$
current and no external electromagnetic field following mostly the conventions of \cite{HoultKovtun}.
We introduce the timelike fluid velocity $u^\mu$ normalised such that $u^2=-1$, and define the operators $\Delta^{\mu\nu}=g^{\mu\nu}+u^\mu u^\nu$,
$D=u^\mu\nabla_\mu$, $\Theta=\nabla_\mu u^\mu$,
$a^\mu=Du^\mu$, $\perpgrad^\mu=\Delta^{\mu\nu}\nabla_\nu$, and
$\alpha=\mu/T$ as well as the shear tensor
\begin{equation}
 \sigma^{\mu\nu}=2\Delta^{\mu\rho}\Delta^{\nu\lambda}
 \nabla_{(\rho}u_{\lambda)}-\frac23\Delta^{\mu\nu}\Theta .
\end{equation}
The conservation equations determining the evolution of the fluid variables are $\nabla_\mu T^{\mu\nu}=0$ and
$\nabla_\mu J^\mu=0$, where the most general stress tensor and charge current under the above assumptions are, respectively
\begin{align}
 T^{\mu\nu}&=(\epsilon+\mathcal E)u^\mu u^\nu
 +(p+\mathcal P)\Delta^{\mu\nu}
 +2u^{(\mu}\mathcal Q^{\nu)}-\eta\sigma^{\mu\nu},\label{eq:stress}\\
 J^\mu&=(n+\mathcal N)u^\mu+\mathcal J^\mu,\label{eq:current}\\
 \begin{pmatrix}\mathcal E\\\mathcal P\\\mathcal N\end{pmatrix}
 &=\begin{pmatrix}\eps_1&\eps_2&\eps_3\\
 \pi_1&\pi_2&\pi_3\\\nu_1&\nu_2&\nu_3\end{pmatrix}
 \begin{pmatrix}D\ln T\\\Theta\\D\alpha\end{pmatrix},\label{eq:scalars}\\
 \mathcal Q^\mu&=\theta_1a^\mu+\theta_2\perpgrad^\mu\ln T
                         +\theta_3\perpgrad^\mu\alpha,\nonumber\\
 \mathcal J^\mu&=\gamma_1a^\mu+\gamma_2\perpgrad^\mu\ln T
                         +\gamma_3\perpgrad^\mu\alpha.\label{eq:vectors}
\end{align}
All transport coefficients are functions of the temperature $T$ and chemical potential $\mu$. In the case of neutral fluids they are functions of  $T$ only and six non-trivial coefficients, $\epsilon_{1,2},\pi_{1,2},\theta_{1,2}$, remain while the others vanish. In contrast to the thermodynamic-frame choices used in
Refs.~\cite{BDN2018,Kovtun2019,BDN2019,HoultKovtun,BDN2022},
we do not impose $\theta_1=\theta_2$ or, for charged fluids,
$\gamma_1=\gamma_2$.
We allow these vector coefficients to differ, subject to the entropy
constraint stated below. This enlarges the frame choices without changing
the equilibrium equation of state.

We are only interested in thermodynamically consistent equations of state (EoS) defined via the pressure $p(T,\mu)$ satisfying $\epsilon=-p+Tp_T+\mu p_\mu$ where we introduced the charge density $n=p_\mu=\partial p/\partial\mu$, the entropy density $s=p_T=\partial p/\partial T$ and the energy density $\epsilon$. To specify the constraints that some of these functions must obey, it is useful to define
\begin{equation}\label{eq:thermo}
 r=\frac n w,\qquad A=\left.\frac{\partial p}{\partial\epsilon}\right|_n,
 \qquad TB=\left.\frac{\partial p}{\partial n}\right|_\epsilon,
 \qquad x=c_s^2=A+rTB~,
\end{equation}
where $w=\epsilon+p$ is the enthalpy and $x$ is the square of the sound speed. We also introduce
\begin{equation}\label{eq:suscept}
 \begin{gathered}
 E=\left.\partial_T\epsilon\right|_\alpha,\qquad
 F=\left.\partial_\alpha\epsilon\right|_T,\qquad
 N=\left.\partial_T n\right|_\alpha,\qquad
 H=\left.\partial_\alpha n\right|_T,\\[3pt]
 \Delta_J=EH-FN=T(\epsilon_Tn_\mu-\epsilon_\mu n_T),\qquad
 q_T=N-rE,\qquad q_\alpha=H-rF~,
 \end{gathered}
\end{equation}
where the subscript $T$ (or $\mu$) denotes a thermodynamic derivative with respect to $T$ (or $\mu$). Given these functions, we work on a regular and causal thermodynamic domain $T>0$ and $w>0$ with the constraints
\begin{equation}\label{eq:domain}
\begin{split}
 &T>0,\quad w>0,\quad\Delta_J>0,\quad 0<x<1,\\
 &n_\mu\geq0,\quad E\geq0,\quad
 A+\frac{n_\mu}{w}(TB)^2\geq rTB,\quad
 n_\mu\geq\frac{n^2}{wx}~.
 \end{split}
\end{equation}
The strict inequality in $\Delta_J$ and $x$ avoid singular susceptibilities and degenerate or luminal sound speeds. Five of these conditions follow from equilibrium fluctuation positivity \cite{HoultKovtun} and their derivation is recalled in Appendix~\ref{app:assumptions}. In addition, the second law of thermodynamics enforces \cite{HoultKovtun}
\begin{equation}\label{eq:thermo-domain}
\eta,\zeta,\cD\geq0,\quad  \gamma_2-\gamma_1=r(\theta_2-\theta_1), 
\end{equation}
as we re-derive in Section~\ref{app:sec:charged-uplift}, where $\eta$ is the shear viscosity, $\zeta$ is the bulk viscosity and $\cD$ the conductivity defined as 
\begin{align}
 \zeta={}&-\pi_2-\eps_1A^2-T\nu_3B^2+A(\eps_2+\pi_1)
       +B(\pi_3+T\nu_2)-AB(\eps_3+T\nu_1),\nonumber\\
 \cD T={}&r(\theta_3-rT\theta_1)+rT\gamma_1-\gamma_3.\label{eq:matching}
\end{align}
We consider the EoS $p(T,\mu)$ and the transport coefficients $\eta,\zeta,\cD$ as the physical data characterising the fluid, and derivable for instance via kinetic theory, holography, or any kind of microscopic model. Our goal is to find families of frames, i.e. choices of the remaining coefficients, in terms of this physical data, satisfying \eqref{eq:domain}-\eqref{eq:thermo-domain} but otherwise generic, that render the BDNK theory linearly causal and stable following the strategy described in the next subsection. 

\subsection{Frame construction strategy}\label{sec:construction-strategy}
We begin by linearising the conservation equations $\nabla_\mu T^{\mu\nu}=0$ and
$\nabla_\mu J^\mu=0$ in 3+1 dimensions around a uniform equilibrium state $T\to T+\delta T$, $\mu\to\mu+\delta \mu$ and $u^\mu\to u^\mu+\delta u^\mu$ in flat space $g_{\mu\nu}=\eta_{\mu\nu}$ using plane-wave
perturbations of the form $e^{-i\omega t+i\mathbf k\cdot\mathbf x}$ where $\omega$ is the frequency and $\bold k$ the wavevector.
The frequencies are given by the roots of a dispersion polynomial $P(\varsigma,\mathbf k;\mathbf v)$. Up to a nonzero
normalization, rotational symmetry in the fluid rest frame and boost invariance \cite{Kovtun2019,HoultKovtun} gives
\begin{equation}\label{eq:strategy-full-polynomial}
 P(\varsigma,\mathbf k;\mathbf v)
 \propto P_T(\varsigma,\mathbf k;\mathbf v)^2P_L(\varsigma,\mathbf k;\mathbf v)=0,
 \qquad \varsigma=-i\omega~.
\end{equation}
Here $\mathbf v$ is the equilibrium spatial fluid velocity which can be generated by a boost from the rest state, $\varsigma$ is the growth rate,
$P_T$ is the polynomial factor for each of the two transverse velocity
polarizations, and $P_L$ describes the longitudinal velocity that couples
to temperature and, for charged fluids, to the chemical potential. Our goal is to find dispersion relations $\omega_a(\mathbf k)$ that are linearly stable and causal as we now explain.

\subsubsection{What the roots must satisfy}
For every root $\omega_a$, every real wavevector, and every
uniform velocity $|\mathbf v|<1$, linear spectral stability requires
\begin{equation}\label{eq:strategy-spectral-condition}
 \operatorname{Im}\omega_a(\mathbf k;\mathbf v)
 =\operatorname{Re}\varsigma_a(\mathbf k;\mathbf v)\leq0.
\end{equation}
This bound must be satisfied for all $\mathbf k$ including $\mathbf k=0$. In order to turn stability into algebraically tractable inequalities, we first work at rest $\mathbf v=0$ and 
define $q=|\mathbf k|^2$. For any real coefficient factor
$p_m(\varsigma,q)=\sum_{j=0}^{m}a_j(q)\varsigma^j$, normalized with $a_m(q)>0$,
the Routh-Hurwitz criterion says that all roots have $\Re \varsigma<0$ if and
only if all its Hurwitz determinants are positive:
\begin{equation}\label{eq:strategy-rh-definition}
 \Delta_j(q)=\det\!\left[a_{m+i-2\ell}(q)\right]_{i,\ell=1}^{j}>0,
 \qquad j=1,\ldots,m.
\end{equation}
Coefficients outside $0\leq j\leq m$ are set to zero. In particular,
$\Delta_1=a_{m-1}$ and
$\Delta_2=a_{m-1}a_{m-2}-a_ma_{m-3}$. We impose their positivity for every $q>0$ while
zeros at $q=0$ are factored out before testing the remaining roots. At nonzero boosts $\mathbf v\ne 0$ the coefficients generally
become complex, so this real coefficient test is supplemented by the direct boosted proofs in Appendix~\ref{app:arbitrary-boosts}. Causality  in turn additionally requires real,
finite characteristic speeds not exceeding the speed of light. These are the large $\mathbf k$ slopes of the dispersion relations. In particular for $\mathbf k=k\hat{\mathbf k}$,
\begin{equation}\label{eq:strategy-characteristic-condition}
 \lim_{k\to\infty}\frac{\omega_a(k\hat{\mathbf k};\mathbf v)}{k}
 =c_a(\hat{\mathbf k};\mathbf v)\in\mathbb R,
 \qquad |c_a|\leq1.
\end{equation}
The speed bound is therefore $\lim_{k\to\infty}|\omega_a|/k\leq1$.
We apply it to the principal part of the full local conservation system.
A bound on an isolated dispersion branch is not, by itself, a causality
proof~\cite{HoultKovtunDispersion,GavassinoDisconziNoronha}.
The polynomial criterion also requires that the degree in frequency
match the total derivative degree, so that no additional modes arise
from spatial derivatives after a boost~\cite{HoultKovtunDispersion}.
Below we check this for the full neutral and charged determinants,
and Appendix~\ref{app:arbitrary-boosts} verifies their characteristic
speeds for every propagation angle.

\subsubsection{Transverse sector: the telegrapher equation}\label{sec:strategy-shear}
The rest frame transverse polynomial is the same for neutral and
charged fluids, and given by a simple quadratic polynomial of the form
\begin{equation}\label{eq:strategy-shear-polynomial}
 P_T(\varsigma,q)=\theta_1\varsigma^2+w\varsigma+\eta q.
\end{equation}
Indeed, a transverse perturbation has $\delta T=\delta\alpha=0$ and
$\mathbf k\cdot\delta\mathbf u=0$, leaving only the acceleration
coefficient $\theta_1$, enthalpy $w$, and shear viscosity $\eta$ in the linear analysis.
Dividing the polynomial by $w$ gives the Fourier polynomial of the
\emph{telegrapher equation} for a transverse velocity amplitude $V_\perp$:
\begin{equation}\label{eq:strategy-telegrapher}
 \tau_T\partial_t^2V_\perp+\partial_tV_\perp-\nu_T\nabla^2V_\perp=0,
 \qquad \tau_T=\frac{\theta_1}{w},\quad \nu_T=\frac{\eta}{w}.
\end{equation}
This is the diffusion equation with an additional second time derivative term. The telegrapher form, which we exploit further in the charged case, makes it straightforward to derive  stable and causal bounds on the BDNK coefficients as we now explain. For
$\tau_T,\nu_T>0$ and at small $q$, it has one diffusive root
$\varsigma=-\nu_Tq+O(q^2)$ and one relaxation root tending to $-1/\tau_T$ as $q\to0$.
The quadratic Routh-Hurwitz test is simply positivity of its coefficients while
its squared characteristic speed is $y_T=\nu_T/\tau_T=\eta/\theta_1$.
Consequently the stability and causality requirements \eqref{eq:strategy-rh-definition} and \eqref{eq:strategy-characteristic-condition} reduce to
\begin{equation}\label{eq:strategy-shear-bound}
 \theta_1\geq\eta>0,
 \qquad \theta_1=\alpha_T\eta,\quad\alpha_T\geq1.
\end{equation}
The choice $\alpha_T>1$ makes this shear channel strictly subluminal. The direct
boosted calculation shows that the same bound controls the shear channel at all
momenta and boosts, including $\alpha_T=1$ (see
Appendix~\ref{app:sec:shear}). We note that it does not require a thermodynamic frame equality between $\theta_1$ and
$\theta_2$. The coefficient $\theta_1$ also enters the longitudinal sector, so its bound must be solved together with the conditions on the longitudinal (sound) channel.

\subsubsection{Neutral longitudinal sector: reducing the quartic}
In the case of neutral fluids, the longitudinal polynomial is a function of the six frame coefficients
$\varepsilon_{1,2}$, $\pi_{1,2}$ and $\theta_{1,2}$.
A choice of these coefficients will also serve as a \emph{seed} for
the charged construction that follows later. We denote its bulk viscosity invariant
by $\zeta_*$, i.e. for neutral fluids $\zeta_*=\zeta$. We also introduce
$\Gamma=4\eta/3+\zeta_*$, which is the combination controlling sound
attenuation. The second law \eqref{eq:thermo-domain} gives $\Gamma\geq0$,
with $\Gamma>0$ in the positive-shear regime considered here. In this case, the longitudinal polynomial $P_L=P_N$ is given by the quartic
\begin{equation}\label{eq:neutral-sound-polynomial}
 P_N=A_L\varsigma^4+B_L\varsigma^3+(w^2+C_Lq)\varsigma^2+D_Lq\varsigma+xw^2q+F_Lq^2~,
\end{equation}
with coefficients in the form
\begin{align}
 A_L&=x\varepsilon_1\theta_1, \quad B_L=w(x\varepsilon_1+\theta_1), \quad C_L=x\{\varepsilon_1(4\eta/3-\pi_2)+\varepsilon_2(\theta_2+\pi_1)+\theta_1\pi_1\},\nonumber\\
 D_L&=xB_L+w\Gamma,\quad F_L=x\theta_2(\pi_2-4\eta/3),\label{eq:quartic-data}\\
\zeta_*&= -\pi_2-x^2\varepsilon_1+x(\varepsilon_2+\pi_1). \nonumber
\end{align}
The last relation is the neutral specialization of \eqref{eq:matching}.
Before applying Routh-Hurwitz, we check the zero-momentum modes:
\begin{equation}\label{eq:neutral-rest-gaps}
 P_N(\varsigma,0)=\varsigma^2(x\varepsilon_1\varsigma+w)
                                  (\theta_1\varsigma+w).
\end{equation}
For nonzero $\varepsilon_1,\theta_1$, this gives two gapless modes and the
gaps $-w/(x\varepsilon_1)$ and $-w/\theta_1$. Since $w,x>0$, damping of
both gaps requires $\varepsilon_1,\theta_1>0$.
We work in this nondegenerate regime and introduce the dimensionless combinations
\begin{equation}\label{eq:invariants}
 R=\frac{D_L}{B_L}=x+\frac{\Gamma}{x\varepsilon_1+\theta_1},\qquad
 \Sigma=\frac{C_L}{A_L},\qquad\Phi=\frac{F_L}{A_L},\qquad
 \Lh=R(\Sigma-R)-\Phi~,
\end{equation}
in terms of which the Hurwitz determinants defined in \eqref{eq:strategy-rh-definition} gives
\begin{align}\label{eq:rh}
 \Delta_1&=B_L,\qquad
 \Delta_4=(xw^2q+F_Lq^2)\Delta_3,\nonumber\\
 \Delta_2&=B_Lw^2+A_LB_L(\Sigma-R)q,\nonumber\\
 \Delta_3&=B_Lw^3\Gamma q+A_LB_L^2\Lh q^2.
\end{align}
The requirement at every momentum has thus become a sign problem for
the coefficients of these low-degree polynomials in $q$.
For causality, substituting $\varsigma=-ick$, $q=k^2$ into the terms of
total derivative order four gives $y^2-\Sigma y+\Phi=0$, where
$y=c^2$. The squared longitudinal and transverse speeds are
\begin{equation}\label{eq:speeds}
 y_\pm=\frac{\Sigma\pm\sqrt{\Sigma^2-4\Phi}}2,
 \qquad y_T=\frac{\eta}{\theta_1}.
\end{equation}
A sufficient set of rest frame design inequalities is therefore
\begin{equation}\label{eq:certificate}
 \Phi\geq0,\qquad \Lh\geq0,\qquad
 0\leq y_-\leq y_+\leq1,\qquad \theta_1\geq\eta.
\end{equation}
Indeed, $\Lh\geq0$ places $R$ between $y_-$ and $y_+$, so
$\Sigma\geq R$ and all four Hurwitz determinants are strictly
positive at every $q>0$. The remaining task is to choose actual
frame coefficients satisfying these inequalities and entropy matching $\zeta_*=\zeta$. In summary, the Routh-Hurwitz inequalities and causality condition for a quartic polynomial are still analytically tractable - a realisation that we will be exploiting in the charged case. 

\subsubsection{Charged longitudinal sector: extracting a telegrapher factor}
For charged fluids, $P_L=P_{\rm ch}$ comes from three coupled
longitudinal equations and is generically a sextic:
\begin{equation}\label{eq:strategy-charged-sextic}
 P_{\rm ch}(\varsigma,q)=\sum_{j=0}^{6}a_j(q)\varsigma^j.
\end{equation}
Directly imposing its six Routh-Hurwitz inequalities for all $q>0$ is
considerably harder to solve explicitly for arbitrary physical inputs. We instead seek frame coefficients for
which the original determinant factorizes exactly as
\begin{equation}\label{eq:strategy-factorization}
 P_{\rm ch}(\varsigma,q)=\mathcal N_0F_q(\varsigma,q)P_N(\varsigma,q),\qquad
 F_q(\varsigma,q)=\tau_q\varsigma^2+\varsigma+D_qq.
\end{equation}
Here $\mathcal N_0\ne0$ is an overall normalisation that depends on thermodynamic data, $D_q$ is
a charge diffusivity, and $\tau_q$ is its relaxation time to be defined precisely below. The factor
$F_q$ has precisely the telegrapher form introduced in the shear
sector in \eqref{eq:strategy-shear-polynomial}. With $D_q>0$, choosing $\tau_q\geq D_q$ gives stable rest
roots and a squared characteristic speed $y_q=D_q/\tau_q\leq1$.
The other factor has the neutral quartic form, evaluated at the charged
values $w,x,\zeta_*$, where
$\zeta_*=\zeta+\mathcal D T^2B^2/x$ as defined in \eqref{eq:effective}.
The charged bulk viscosity remains the prescribed $\zeta$ in
\eqref{eq:matching}. Thus the search reduces to realizing this
factorization while preserving the physical transport coefficients. At rest and $q=0$, $P_{\rm ch}$ gives three gapless and three
gapped modes.\footnote{For positive transport and nonzero leading coefficients, the frequency
and total derivative degrees agree: eight for $P_T^2P_N$ and ten for
$P_T^2F_qP_N$, as required by the polynomial causality
criterion~\cite{HoultKovtunDispersion}. Appendix~\ref{app:arbitrary-boosts}
verifies that these degrees are unchanged under finite boosts.
The equality of frequency and total derivative degrees also holds when
certain transport coefficients vanish.}

\subsubsection{Five step frame construction strategy} \label{sec:constructionstrategy}
Below we detail the strategy used to find families of BDNK frames for
any given EoS and transport coefficients.
\begin{enumerate}
\item \textbf{Solve the entropy relations.}
The first step is to solve the entropy relations \eqref{eq:matching} for some of the BDNK coefficients in terms of the physical coefficients $\cD, \zeta$ and the remaining BDNK coefficients.
In the neutral case, for instance, we solve the last relation in
\eqref{eq:quartic-data}, for example as
$\pi_1=x\varepsilon_1+(\zeta_*+\pi_2)/x-\varepsilon_2$. The remaining coefficients will be fixed in the later steps.

\item \textbf{Choose branches that make the reduced quartic inequalities solvable.}
Next, we compute $R,\Sigma,\Phi,\Lh$ using the coefficients in the previous step and impose
\eqref{eq:certificate}. To simplify the inequalities, we set $\pi_2=4\eta/3$ in order to obtain $\Phi=0$, which reduces the sound bounds to
$R\leq\Sigma\leq1$. Alternatively taking $\varepsilon_2=0$ extracts the exact factor
$\theta_1\varsigma+w$ from the polynomial, leaving a simpler cubic polynomial to analyse. Keeping $\varepsilon_2\ne0$
and choosing $\Sigma=1$ leaves room for families that admit conformal limits. Other simplifying alternatives are possible and we discuss them in Section~\ref{sec:factorization}.

\item \textbf{Optional: impose a finite conformal limit.}
Certain EoS and transport coefficients may admit a conformal point. At
that point, conformal symmetry requires finite coefficients satisfying
\cite{Kovtun2019,HoultKovtun}
\begin{equation}\label{eq:conformal}
 \eps_i=3\pi_i,\quad \pi_1=3\pi_2,\quad \nu_1=3\nu_2,
 \quad\theta_1=\theta_2,\quad\gamma_1=\gamma_2.
\end{equation}
In the neutral case, these constraints are equivalently expressed by requiring the seed
to approach 
\[
 (\varepsilon_1,\varepsilon_2,\pi_1,\pi_2,\theta_1,\theta_2)
 \longrightarrow(3\Pi,\Pi,\Pi,\Pi/3,\theta,\theta)
\]
as $x\to1/3$ and $\zeta_*\to0$, with finite $\Pi,\theta$. This means that non-thermodynamic frame choices are allowed away
from the conformal point, but $\theta_2-\theta_1$ must vanish at the conformal point. 

\item \textbf{Uplift each neutral seed to the charged factorization.}
Each admissible neutral seed can be \emph{uplifted} to a charged family by
choosing the additional coefficients so that, after linearising about
uniform equilibrium and Fourier transforming in its rest frame,
$\delta[\nabla_\mu J^\mu+r u_\nu\nabla_\mu T^{\mu\nu}]
=F_q(\varsigma,q)h=0$.
Here $h=\delta n-r\delta\epsilon=q_T\delta T+q_\alpha\delta\alpha$,
and $r=n/w$ is evaluated at equilibrium. The coefficient map in
Section~\ref{sec:charged-factorization-overview} realizes
\eqref{eq:strategy-factorization}, with
\begin{equation}\label{eq:effective}
 \begin{gathered}
 \Xi=q_\alpha-rT^2q_T=\frac{T\Delta_Jx}{w}>0,\qquad
 D_q=\frac{\cD T}{\Xi},\qquad
 \tau_q=\alpha_qD_q,\quad\alpha_q\geq1,\\[3pt]
 \zeta_* =\zeta+\frac{\cD T^2B^2}{x}.
 \end{gathered}
\end{equation}
Here $\Xi$ is a charge susceptibility and $\alpha_q$ is a free
dimensionless relaxation parameter. Entropy matching \eqref{eq:matching}
requires evaluating the neutral seed at the charged values $w,x,\zeta_*$.
The physical charged bulk viscosity remains the prescribed $\zeta$.

\item \textbf{Verify the original determinant, boosts and boundaries.}
We substitute the proposed coefficients into the constitutive relations to verify \eqref{eq:matching} with the prescribed transport coefficients and compute the determinant of the resulting linearized conservation equations to verify the factorization of the polynomial exactly. We then use the boosted polynomial
at real momentum, including its actual $\mathbf k=0$
gaps to make sure that the boosted modes are stable.  In practice, in Appendix~\ref{app:arbitrary-boosts} we show that the reduced quartic inequalities and telegrapher bounds already ensure stability and causality in every inertial frame. Hence no additional coefficient inequalities arise from boosting, and the result applies to every family of frames found by following the previous steps above. When frame coefficients or physical transport coefficients vanish, the
polynomials can change degree. We then check the reduced equations
directly, as summarized in footnotes~\ref{fn:neutral-zero-transport},
\ref{fn:shear-zero-transport} and \ref{fn:charged-zero-transport}.
\end{enumerate}
The steps outlined above, together with brute-force polynomial
factorisation, form the basis of the
\href{https://jacomearmas.org/research/BDNK}{AI research protocol}. 
We will now apply this strategy to neutral and charged fluids.

\subsection{Neutral construction and charged telegrapher factorization}\label{sec:factorization}
We apply the strategy of Section~\ref{sec:construction-strategy} to
construct the neutral families \family{1}--\family{3}, \family{4}, \family{5} and \family{6}, and then uplift
them to the charged case. \family{4}, \family{5} and \family{6} admit finite conformal limits
with the matching conditions stated below.

\subsubsection{Neutral branches without conformal limits}
\label{sec:neutral-branches-overview}
We begin by solving the reduced conditions \eqref{eq:certificate} by choosing the
remaining frame coefficients after entropy matching \eqref{eq:matching}. Setting $\pi_2=4\eta/3$ makes $\Phi=0$ and reduces the sound
constraints to the interval $R\leq\Sigma\leq1$. Setting $\varepsilon_2=0$ yields, for any $\theta_2$, the exact factorization
\begin{equation}\label{eq:relax-factor}
 P_N=(\theta_1\varsigma+w)\bigl[x\varepsilon_1\varsigma^3+w\varsigma^2+x\pi_1q\varsigma+xwq\bigr].
\end{equation}
Equation~\eqref{eq:relax-factor} is an exact simplification of the neutral sound polynomial:
it separates the momentum-independent relaxation root $\varsigma=-w/\theta_1$
(when $\theta_1>0$) from a cubic polynomial describing two sound modes and a second relaxation mode.
Using entropy matching \eqref{eq:matching}, the cubic part of the polynomial becomes
\[
 (w+x\varepsilon_1\varsigma)(\varsigma^2+xq)+\Gamma \varsigma q.
\]
The term $\Gamma \varsigma q$ couples sound to relaxation.\footnote{\label{fn:neutral-quadratic-obstruction}One can show that the full
neutral sound quartic polynomial cannot factor exactly into two local scalar quadratic
polynomials in $(\varsigma,k)$, with coefficients depending only on state
and transport data, for a family with a finite, stable and causal conformal
limit and $\eta_c>0$. Matching coefficients in this factorization ansatz,
together with the conformal relations, forces
$\eta_c(2\Pi+\eta_c)=0$, whereas stability requires $\Pi\geq4\eta_c>0$.
This obstruction concerns the stated local quadratic ansatz. It does
not exclude factorizations with non-polynomial momentum dependence.} Its cubic
Routh-Hurwitz inequality reduces to positivity of $\Gamma$ which is already guaranteed by the second law \eqref{eq:thermo-domain} while
causality fixes $\varepsilon_1\geq\Gamma/[x(1-x)]$. Together with $\theta_1\geq\eta$ from \eqref{eq:strategy-shear-bound},
this defines the families \family{1} and \family{2} by choosing $\theta_2=0$ (non-thermodynamic) and $\theta_2=\theta_1$ (thermodynamic), respectively.
The family \family{1} allows equality in both causality bounds, $\varepsilon_1=\Gamma/[x(1-x)]$ and $\theta_1=\eta$.

Alternatively, we can retain $\varepsilon_2\ne0$ and impose $\Sigma=1$, thereby determining
\begin{equation}\label{eq:luminal-design}
 \theta_2=\frac{(\varepsilon_1-\pi_1)\theta_1-\varepsilon_2\pi_1}{\varepsilon_2},\qquad
 x\varepsilon_1+\theta_1\geq\frac{\Gamma}{1-x}.
\end{equation}
This defines a general luminal branch: $\varepsilon_1$, $\varepsilon_2\ne0$
and $\theta_1$ remain free, subject to $\varepsilon_1>0$,
$\theta_1\geq\eta$ and the inequality in \eqref{eq:luminal-design}.
Entropy matching fixes $\pi_1=x\varepsilon_1+\Gamma/x-\varepsilon_2$,
and \eqref{eq:luminal-design} fixes $\theta_2$.
The particular choices $\varepsilon_2=-2\eta$ and $\theta_1=\eta$
give rise to the family that we refer to as \family{3} and its remaining bound on $\varepsilon_1$ below.
In summary, the three prescriptions just described are
\begin{align}
 \family{1},\family{2}:\quad &(\varepsilon_1,\varepsilon_2,\pi_1,\pi_2,\theta_1)
       =(e,0,xe+\Gamma/x,4\eta/3,\alpha_T\eta),
       \quad e=\frac{\lambda\Gamma}{x(1-x)},\nonumber\\
 &\theta_2=0\ \text{(\family{1})},\qquad \theta_2=\theta_1\ \text{(\family{2})},
       \quad\lambda,\alpha_T\geq1,\nonumber\\
 \family{3}:\quad &\varepsilon_2=-2\eta,\quad \pi_1=x\varepsilon_1+\Gamma/x+2\eta,
       \quad \pi_2=4\eta/3,\quad \theta_1=\eta,\nonumber\\
 &\theta_2=-(\varepsilon_1+\pi_1)/2,\qquad
       \varepsilon_1\geq\frac{\Gamma-\eta(1-x)}{x(1-x)}.\label{eq:S-seeds}
\end{align}
These bounds ensure $\varepsilon_1,\theta_1>0$, so the gapped modes
in \eqref{eq:neutral-rest-gaps} are damped. For \family{3}, positivity
follows from $\Gamma-\eta(1-x)=\zeta_*+\eta(x+1/3)>0$.
Here $\lambda$ increases the scalar relaxation coefficient and
$\alpha_T$ increases the shear relaxation time relative to their causal
lower bounds and can be chosen freely. At $\lambda=\alpha_T=1$, the family \family{1} has
$\pi_1=\varepsilon_1$. Substituting its coefficients \eqref{eq:S-seeds}
into the characteristic-speed formulas \eqref{eq:speeds} gives
$y_-=0$ and $y_+=y_T=1$: one longitudinal characteristic speed vanishes,
while the other longitudinal speed and the shear speed equal the speed of light.
For positive limiting shear viscosity, these families do not admit
conformal limits, as shown in Section~\ref{app:sec:neutral-family-derivations}.
We now return to the general branch to construct families with conformal limits.

\subsubsection{Neutral families with finite conformal limits}
\label{sec:conformal-construction}
We now construct the neutral families \family{4}, \family{5} and \family{6} by solving the reduced inequalities and then
imposing the finite conformal limits required by \eqref{eq:conformal}. To construct \family{4}, we return to the general branch \eqref{eq:luminal-design}
with $\pi_2=4\eta/3$ and $\Sigma=1$, before imposing \family{3}'s choices
$\varepsilon_2=-2\eta$ and $\theta_1=\eta$.
A well defined conformal limit requires $\varepsilon_2\to3\pi_2=4\eta_c>0$
when the limiting shear viscosity $\eta_c$ is positive. We therefore choose $\varepsilon_2$ and $\pi_1$ both positive and parametrise their freedom in the following way.
Entropy matching \eqref{eq:matching}  fixes only the sum
$\varepsilon_2+\pi_1=x\varepsilon_1+\Gamma/x$ so we can parametrise the remaining freedom by
$\varepsilon_1=L\Gamma$, $\theta_1=K\Gamma$ for some $L$, and a fraction $f$ of
that sum assigned to $\pi_1$. The luminal condition
\eqref{eq:luminal-design} then fixes $\theta_2$, giving
\begin{align}
 \varepsilon_1&=L\Gamma,\quad\varepsilon_2=(1-f)S_0\Gamma,\quad \pi_1=fS_0\Gamma, \quad S_0=xL+1/x,\nonumber\\
 \pi_2&=4\eta/3,\theta_1=K\Gamma,\quad  \theta_2=\Gamma\left[\frac{K(L-fS_0)}{(1-f)S_0}-fS_0\right].\label{eq:AG}
\end{align}
The choices
$L\geq2/[x(1-x)]$, $K\geq3/4$ and $0<f<1$ solve the remaining bounds as they
 give $R\leq(1+x)/2<1$, $\theta_1\geq\eta$ and a nonzero
denominator in $\theta_2$. Thus $\Phi=0$, $\Sigma=1$ and
$\Lh=R(1-R)>0$ satisfying \eqref{eq:certificate}. These bounds establish stability and causality but not yet a conformal limit. In order to do so we assume that  $\eta_c>0$ denotes the limiting conformal form of the shear viscosity. Since
$\pi_2\to4\eta_c/3$, Step~3 in Section~\ref{sec:constructionstrategy} fixes $\Pi=4\eta_c$.
The limits $\varepsilon_1\to3\Pi$ and
$\varepsilon_2,\pi_1\to\Pi$ therefore require $L\to9$ and $f\to1/2$.
Using these scalar limits and $\theta_1\to\theta$, the luminal
condition \eqref{eq:luminal-design} gives $\theta_2\to2\theta-4\eta_c$. Requiring
$\theta_2\to\theta$ fixes $\theta=4\eta_c$, hence $K\to3$.
These implement the conformal restrictions leading to a family \family{4} with a well defined conformal limit and is in general a non-thermodynamic frame.\footnote{\label{fn:F4-existing-subset}The subset $f=1/(1+x^2L)$ and $K=L/[(1-x)L-1/x]$ belongs to the neutral class of Ref.~\cite{BDN2022}: it obeys $\varepsilon_2=x\varepsilon_1$, $\pi_2=x\pi_1-\zeta$ and $\theta_1=\theta_2$, with $\tau_\epsilon=xL\Gamma/w$, $\tau_P=\Gamma/w$ and $\tau_Q=K\Gamma/w$. The bound on $L$ makes the denominator positive and implies $K>1/(1-x)>1$. At conformality, $L\to9$ gives $f\to1/2$ and $K\to3$ (see Appendix~\ref{app:existing-neutral}).}

In turn, to construct the family \family{5} with a conformal limit, we first pick the \family{4} scalar parameters to be
$L=2/[x(1-x)]$ and $f=1/2$, which satisfy the preceding bounds and
conformal restrictions and imply
$m=\varepsilon_2=\pi_1$. Contrary to the family \family{4}, we now impose a thermodynamic frame by choosing $\theta_1=\theta_2=\theta_0$ and hence \eqref{eq:luminal-design} gives
$(\varepsilon_1-2m)\theta_0=m^2$. This determines the parameters
\begin{equation}\label{eq:balanced}
 \varepsilon_1=\frac{2\Gamma}{x(1-x)},\quad
 m=\frac{\Gamma(1+x)}{2x(1-x)},\quad \pi_2=\frac43\eta,
 \quad \theta_0=\frac{\Gamma(1+x)^2}{4x(1-x)^2}.
\end{equation}
Keeping these parameters as they are, we can deform away from the luminal branch $\Phi=0,\Sigma=1$ by increasing slightly the coefficients $\theta_1,\theta_2$ such that
$\theta_1=\theta_2=\lambda\theta_0$, with $\lambda>1$. In this case the condition $\Phi=0$ remains unchanged, while
\[
 \Sigma=\rho_0=\frac{1+x}{2}+\frac{1-x}{2\lambda}<1.
\]
For these choices one finds that $R<\rho_0$, and
$\lambda\theta_0>\eta$ also makes the shear speed \eqref{eq:speeds} strictly subluminal. It is possible to slightly generalise this family to non-thermodynamic frames by keeping
$\theta_1=\lambda\theta_0$ and choosing
$\theta_2=\theta_1+\chi(\lambda-1)m$, where
\begin{equation}\label{eq:chi}
 \lambda>1,\qquad
 \chi=\beta\frac{(3x-1)^2}{1+(3x-1)^2},\qquad0\leq\beta\leq1.
\end{equation}
Here the parameter $\beta$ parametrises the freedom in choosing $\chi$. This changes the upper squared sound speed $y_+$ \eqref{eq:speeds} by a fraction $\chi<1$ of
the available margin to the light cone:
\[
 y_-=0,\qquad y_+=\rho=\rho_0+\chi(1-\rho_0)
 =1-\frac{(1-x)(\lambda-1)(1-\chi)}{2\lambda}<1.
\]
Hence $R<\rho_0\leq\rho<1$ preserves the reduced stability and
causality inequalities. The explicit positive margins are proved in
Section~\ref{app:sec:neutral-conformal-derivations}.
Since $\chi\to0$ at conformality, finite parameters with
$\lambda\to\lambda_c>1$ give the Step~3 limits in Section~\ref{sec:constructionstrategy}
$\Pi=4\eta_c$ and $\theta=4\lambda_c\eta_c$.
Thus \family{5} achieves strict subluminality without losing its conformal limit while
its lower sound speed $y_-$ remains zero.

The family \family{6} that we now describe allows for non-zero lower sound speed. In order to construct it, 
we choose $\pi_2>4\eta/3$ and $\theta_2>0$,
which makes $\Phi>0$ in \eqref{eq:quartic-data}. We introduce a parameter
$\Lambda$ to parametrise this freedom, while the function $\chi$ in
\eqref{eq:chi} controls the difference between the $\theta_1$ and $\theta_2$. Choosing $1<\Lambda\leq2$, $0\leq\beta\leq1$, this leads to
\begin{align}
 \varepsilon_1&=\frac{2\Lambda\Gamma}{x(1-x)},&
 \varepsilon_2=\pi_1&=\frac{\Lambda\Gamma(1+x)}{2x(1-x)},\nonumber\\
 \pi_2&=\frac43\eta+(\Lambda-1)\Gamma,&
 \theta_1&=\frac{\Lambda\Gamma(1+x)^2}{2x(1-x)^2},\nonumber\\
 \theta_2&=\theta_1\left[1+\frac{4\chi(1-\Sigma_0)}{1+x}\right],&
 \Sigma_0&=\frac{3+x}{4}
 -\frac{2(\Lambda-1)x(1-x)^2}{\Lambda(1+x)^2}.
 \label{eq:AN}
\end{align}
The squared sound speed combinations defined in \eqref{eq:invariants} are
then
\begin{align}
 \Sigma&=\Sigma_0+\chi(1-\Sigma_0),&
 \Phi&=\frac{(\Lambda-1)x(1-x)}{2\Lambda}
       +\frac{2(\Lambda-1)x(1-x)}{\Lambda(1+x)}\chi(1-\Sigma_0),
       \nonumber\\
 R&=x+\frac{2x(1-x)^2}{\Lambda(1+6x-3x^2)},\label{eq:AN-speeds}
\end{align}
and satisfy $\Phi>0$, $\Sigma<1$ and $\Lh>0$ given in \eqref{eq:certificate}. The resulting neutral speed bounds are
\begin{equation}\label{eq:AN-bounds}
 0<y_-<R<y_+<1,\qquad 0<y_T<\frac1{2\Lambda}<1.
\end{equation}
The choice
$\beta=0$ yields a thermodynamic frame while $\beta>0$ permits
non-thermodynamic frames away from $x=1/3$. Bounded parameters with
finite conformal limits yield finite coefficients. In particular,
$\Lambda\to\Lambda_c$ and $\chi\to0$ give the Step~3 limits (see Section~\ref{sec:constructionstrategy})
$\Pi=4\Lambda_c\eta_c$ and $\theta=8\Lambda_c\eta_c$.
For $\eta_c>0$ and $1<\Lambda_c\leq2$, both longitudinal speeds
remain nonzero and strictly subluminal at conformality. The full derivation
is given in Section~\ref{app:sec:neutral-conformal-derivations}.

\subsubsection{Charged uplift: an additional exact telegrapher factor}
\label{sec:charged-factorization-overview}
Having constructed the neutral families \family{1}-\family{3}, \family{4}, \family{5} and \family{6}, we now
provide the coefficient map between neutral and charged frames implementing Step~4 in Section~\ref{sec:constructionstrategy}. Assuming $\cD>0$, we take
$D_q,\tau_q,\zeta_*$ from \eqref{eq:effective} and evaluate the chosen
neutral families at the charged values of $w,x,\eta$, with
$\Gamma=4\eta/3+\zeta_*$. In addition we retain the values of the six stress coefficients
$\varepsilon_{1,2},\pi_{1,2},\theta_{1,2}$ as given by each family and set the charge coefficients according to
\begin{align}
 \eps_3&=rT\eps_1,\qquad \pi_3=rT\pi_1,\qquad
 \theta_3=rT\theta_2,\label{eq:uplift-stress}\\
 (\nu_1,\nu_2,\nu_3)&=
 (r\varepsilon_1+T\tau_qq_T,\ r\varepsilon_2+TD_qq_T,\ r^2T\varepsilon_1+\tau_qq_\alpha),\nonumber\\
 (\gamma_1,\gamma_2,\gamma_3)&=
 (r\theta_1-TD_qq_T,\ r\theta_2-TD_qq_T,\ r^2T\theta_2-D_qq_\alpha).
 \label{eq:uplift-current}
\end{align}
These coefficients preserve the entropy identities $\eta,\zeta,\cD$ in
\eqref{eq:matching}, factorise the polynomial into a telegrapher part times the quartic
and fix the normalization in \eqref{eq:strategy-factorization} such that
\begin{equation}\label{eq:factorization}
 \det M_{\rm charged}=-\frac{\Delta_J}{w}F_q(\varsigma,q)P_N(\varsigma,q).
\end{equation}
We verify this factorization explicitly  in
Appendix~\ref{app:sec:charged-uplift} while stability at arbitrary boosts is verified in Appendix~\ref{app:sec:charged-boost-proof}.  With this map, choosing $\alpha_q>1$ makes
the extra squared speed $y_q=1/\alpha_q$ strictly subluminal as described in
Section~\ref{sec:construction-strategy}. For a regular conformal EoS limit with $\zeta\to0$, one has
$B,q_T\to0$ and $\zeta_*\to0$ leading to finite charged coefficients satisfying the conformal requirements \eqref{eq:conformal}.

We note that in the case of the \family{6} family, choosing $\alpha_q>1$, the bounds \eqref{eq:AN-bounds}, together with
$0<y_q=1/\alpha_q<1$, make every comoving characteristic speed nonzero
and strictly subluminal when $\eta,\cD>0$. As a useful example, at the conformal point, picking $\Lambda=\alpha_q=2$ gives
\begin{equation}\label{eq:benchmark}
 (\varepsilon_1,\varepsilon_2,\pi_1,\pi_2,\theta_1,\theta_2)
 =\eta(24,8,8,8/3,16,16),\qquad
 (y_-,y_+,y_T,y_q)=(1/12,2/3,1/16,1/2),
\end{equation}
as well as $q_T=0$ and $\Xi=q_\alpha$ so
\eqref{eq:effective} gives $D_qq_\alpha=\cD T$ and
$\tau_qq_\alpha=2\cD T$. The current coefficients in
\eqref{eq:uplift-current} therefore become
\begin{align}
 (\nu_1,\nu_2,\nu_3)
   &=(24r\eta,\,8r\eta,\,24r^2T\eta+2\cD T),\nonumber\\
 (\gamma_1,\gamma_2,\gamma_3)
   &=(16r\eta,\,16r\eta,\,16r^2T\eta-\cD T).
 \label{eq:benchmark-current}
\end{align}
Here $r=n/w$ and the physical conductivity $\cD$ remains a free parameter, in particular, $\nu_1=3\nu_2$ and $\gamma_1=\gamma_2$
satisfy the conformal current relations in \eqref{eq:conformal}. We overview all families in the next subsection.

\subsection{Catalogue of families of frames and remaining freedom}\label{sec:catalogue}
Table~\ref{tab:families} collects the six families. Each has a neutral
version and a charged uplift; the latter uses $\Gamma$ from
\eqref{eq:quartic-data} and adds the charge speed $y_q$.
Families \family{1}--\family{3} do not have a finite conformal limit
with positive limiting shear viscosity. Families \family{4}--\family{6}
do, subject to the parameter limits specified in
Section~\ref{sec:conformal-construction}.
Some families overlap at limiting parameter values. As $\lambda\downarrow1$,
\family{5} approaches the thermodynamic member of \family{4} with
\begin{equation}\label{eq:AG-equal-vector}
 L=\frac{2}{x(1-x)},\qquad f=\frac12,\qquad
 K=\frac{(1+x)^2}{4x(1-x)^2},
\end{equation}
for which $\theta_1=\theta_2=\theta_0$ and $y_+=1$.
As $\Lambda\downarrow1$, \family{6} approaches the $\lambda=2$ member
of \family{5} with the same $\beta$, while $y_-$ reaches zero.
These relations are proved in Section~\ref{app:sec:neutral-conformal-derivations}.

The catalogue is not exhaustive. Even after fixing the charged telegrapher
factor, some coefficient changes preserve the spectrum and the physical
transport. For arbitrary regular functions $f_E,f_P,f_Q$ of the allowed
thermodynamic and transport data, let $\delta$ denote the additive changes
\begin{equation}\label{eq:free-shifts}
 \begin{split}
 (\delta\eps_1/T,\delta\eps_3)&=f_E(q_T,q_\alpha),\qquad
 (\delta\pi_1/T,\delta\pi_3)=f_P(q_T,q_\alpha),\\
 (\delta\theta_2/T,\delta\theta_3)&=f_Q(q_T,q_\alpha),
 \end{split}
\end{equation}
with $\delta\nu_{1,3}=r\delta\eps_{1,3}$,
$\delta\gamma_{2,3}=r\delta\theta_{2,3}$ and all other coefficient shifts zero.
These changes preserve the full determinant, the shear polynomial and the
entropy invariants. Finite conformal matching requires, in particular,
$f_E-3f_P\to0$. For a fixed neutral representative and fixed charge factor,
these are the general changes within the charge-row construction used
here, as proved in Appendix~\ref{app:sec:family-hierarchy}. This is a
completeness statement for that row identity; it does not classify all
possible ways of obtaining a telegrapher factor in a charged determinant.

\begin{table}[H]
\centering
\small
\setlength{\tabcolsep}{4pt}
\renewcommand{\arraystretch}{1.35}
\begin{tabularx}{\textwidth}{@{}l c c Y Y@{}}
\toprule
Family & $(y_-,y_+)$ & $y_T$ & Finite conformal limit & Type of frame\\
\midrule
\family{1} & $(0,x+(1-x)/\lambda)$ & $1/\alpha_T$ & No & Non-thermodynamic\\
\family{2} & $(0,x+(1-x)/\lambda)$ & $1/\alpha_T$ & No & Thermodynamic\\
\family{3} & $(0,1)$ & $1$ & No & Non-thermodynamic\\
\family{4} & $(0,1)$ & $\eta/(K\Gamma)$ & Yes: $L\to9$, $K\to3$, $f\to1/2$ & Thermodynamic or non-thermodynamic\\
\family{5} & $(0,\rho)$ & $\eta/(\lambda\theta_0)$ & Yes: finite $\lambda\to\lambda_c>1$ & Thermodynamic if $\chi=0$; otherwise non-thermodynamic\\
\family{6} & $0<y_-<y_+<1$ & $0<\eta/\theta_1<1$ & Yes: $\Lambda\to\Lambda_c$, $1<\Lambda_c\leq2$ & Thermodynamic if $\chi=0$; otherwise non-thermodynamic\\
\bottomrule
\end{tabularx}
\caption{Squared comoving characteristic speeds, with $\eta>0$ and, for
charged uplifts, $\mathcal D>0$. Here
$\rho=1-(1-x)(\lambda-1)(1-\chi)/(2\lambda)<1$ in \family{5} while
\family{6} uses \eqref{eq:AN-speeds}. Each charged uplift adds
$y_q=1/\alpha_q$, with finite $\alpha_q\geq1$ (strictly $>1$ for
\family{5} and \family{6}). The shear speed $y_T$ occurs in two transverse polarizations.
The column with the conformal limit assumes a regular endpoint with $\eta_c>0$.
In \family{5} and \family{6}, $\chi$ is defined by \eqref{eq:chi} and
vanishes at conformality even for $\beta>0$. In \family{4}, the
thermodynamic subfamily is selected by $\theta_1=\theta_2$ in
\eqref{eq:AG}, which involves a relation among $L,K,f$. For vanishing
transport coefficients, the reduced modes are described in
footnotes~\ref{fn:neutral-zero-transport}, \ref{fn:shear-zero-transport}
and \ref{fn:charged-zero-transport}.}
\label{tab:families}
\end{table}

\subsection{Discussion and numerical use}\label{sec:discussion}
Table~\ref{tab:families} provides explicit coefficient prescriptions that
can be used in numerical implementations of neutral and charged BDNK
hydrodynamics. Given an equation of state and prescribed
$\eta,\zeta,\mathcal D$, one evaluates the thermodynamic quantities,
checks \eqref{eq:domain}--\eqref{eq:thermo-domain}, and selects frame
parameters satisfying the bounds of the chosen family. The neutral
coefficients and the common charged uplift are then given in
Sections~\ref{sec:neutral-branches-overview}--\ref{sec:charged-factorization-overview}.
The construction avoids solving the full stability inequalities anew
for each equation of state.

The families allow different choices of characteristic structure while
keeping physical transport fixed. With positive shear viscosity and
conductivity, \family{6} has strictly positive, strictly subluminal
squared comoving characteristic speeds, including at its finite conformal
endpoint. It is therefore a natural starting point. Families \family{1}-\family{5} instead
allow a zero sound characteristic speed, and some also allow luminal
propagation. It would be interesting to test these families in numerical implementations.
In particular, various numerical studies in multiple contexts could provide useful tests of these BDNK families \cite{PandyaPretorius, PandyaMostPretorius, PandyaIdealGas, BantilanBeaFigueras, BeaFigueras, BhambureEtAl, FantiniRubio, ClarisseEtAl,  ShumEtAl}.

\label{sec:nonlinear-discussion}
Nonlinear causality and local existence and uniqueness in Gevrey classes
have been established for several neutral constitutive
classes~\cite{BDN2018,DisconziGevrey,BDN2019}. Applying these results
requires matching the constitutive relations as well as checking the
coefficient, regularity and initial-data hypotheses. For the thermodynamic
neutral choices $\theta_1=\theta_2$, the coefficients used in
Refs.~\cite{BDN2019,BDG} are related to ours by
\[
 (\chi_1,\chi_2,\chi_3,\chi_4)
 =(x\varepsilon_1,\varepsilon_2,x\pi_1,\pi_2),\qquad
 \lambda_{\mathrm{BDN}}=\theta_1=\theta_2.
\]
Under this identification, our causal bounds imply the algebraic
conditions of Ref.~\cite{BDN2019}. Consequently, its results on local
existence, uniqueness and nonlinear causality apply to the thermodynamic
neutral members of our families, provided the constitutive functions
and initial data satisfy the regularity and positivity assumptions
stated there.

Under the same coefficient identification, the neutral \family{6}
subfamily with $\beta=0$, constant $1<\Lambda\leq2$ and $\eta>0$
satisfies the algebraic conditions of Ref.~\cite{BDG}. Its result on
local well-posedness in Sobolev spaces therefore applies in Minkowski
spacetime, provided the EoS and constitutive functions satisfy the
analyticity and positivity assumptions stated there, and the initial
data satisfy the required Sobolev regularity and compatibility
conditions.

For charged fluids, Ref.~\cite{BDN2022} establishes nonlinear causality
and local well-posedness in a constitutive class with $J^\mu=nu^\mu$.
Our current contains additional derivative terms, so those results do
not apply directly. Nonlinear causality conditions for generic charged frames, including
non-thermodynamic frames, were derived in
Ref.~\cite{AbboudSperanzaNoronha} from the characteristics of the full
nonlinear equations. All six charged frame families constructed here
satisfy these conditions for $\eta>0$ and $\mathcal D>0$, within their
stated thermodynamic and parameter domains. Strong hyperbolicity has also been established for a restricted class
of charged constitutive relations in Ref.~\cite{SchianchiAbalos}.
Those results do not apply directly to the general families constructed
here. Their result does apply to the exactly conformal family \family{6}
with $\Lambda=\alpha_q=2$, provided the prescribed conductivity
satisfies the additional restriction
$\mathcal D=4\eta\Delta_J/(3w^2)$. Establishing
well-posedness for the full families constructed here remains a further
step beyond their linear stability and characteristic-speed bounds.

A natural extension of the method developed here is magnetohydrodynamics (MHD) for which a BDNK approach has been developed in various works \cite{ArmasCamilloni, LierJainArmasPorth, HoultKovtunMHD, LierArmasPorth}. These developments motivate seeking coefficient families that keep
anisotropic viscosities, resistivities and other physical transport
prescribed. The magnetic field introduces direction dependent
Alfv\'en and magnetosonic sectors and a magnetic constraint, so the
neutral-to-charged map we established cannot simply be reused but the overall strategy of telegrapher factorisation of the longitudinal polynomial may work. The same strategy may also be useful for chiral fluids \cite{AbboudSperanzaNoronha} and for relativistic superfluids \cite{HoultShukla}. It would be interesting to apply the
\href{https://jacomearmas.org/research/BDNK}{AI research protocol} developed through this work
to these more complicated cases.

\subsubsection*{Acknowledgements}
J. Armas is partly funded by the Dutch Institute for Emergent Phenomena
(DIEP) cluster at the University of Amsterdam via the DIEP programme
Foundations and Applications of Emergence (FAEME) and the national
NWA-ORC 2023 consortium EAAS. The author used OpenAI Codex Astra 6 Pro
to assist with manuscript revision, literature cross-checks, and review
of mathematical arguments and analysis code. The author performed human
checks on the AI-assisted material and takes responsibility for the final
content and conclusions.

\appendix
\setcounter{secnumdepth}{3}
\addtocontents{toc}{\protect\setcounter{tocdepth}{2}}
\hypersetup{bookmarksdepth=3}
\section{Derivations, checks and proofs}\label{app:proofs}

This appendix is organized into three parts.
Section~\ref{app:neutral-frames} derives the neutral entropy production,
rest-frame stability and causality conditions, and frame families.
Section~\ref{app:charged-frames} develops the charged uplift, verifies
entropy matching and determinant factorization, and describes the
remaining frame freedom.
Section~\ref{app:arbitrary-boosts} proves stability and causality for
sound, shear and charge modes in arbitrary inertial frames. Throughout, we use the constitutive notation and thermodynamic data of
Section~\ref{sec:constitutive}, the Fourier convention in
\eqref{eq:strategy-full-polynomial}, and the spectral and characteristic
criteria \eqref{eq:strategy-spectral-condition} and
\eqref{eq:strategy-characteristic-condition}. All coefficients in the
linear calculations are evaluated at $(T_0,\mu_0)$, with unsubscripted
state quantities denoting those equilibrium values.

\subsection{Neutral frames}\label{app:neutral-frames}
For a neutral fluid, $dp=s\,dT$, $d\epsilon=T\,ds$ and $w=Ts$ imply
\begin{equation}\label{app:neutral-domain}
 x=\frac{p_T}{\epsilon_T}=\frac{w}{T\epsilon_T}\in(0,1).
\end{equation}
We require $T,w>0$ and the viscosity bounds in \eqref{eq:thermo-domain},
and set $\zeta_*=\zeta$ in the neutral calculations.
The conformal coefficients $\Pi,\theta$ are those introduced in Step~3
of Section~\ref{sec:constructionstrategy}.

\subsubsection{Neutral entropy production}\label{app:neutral-entropy}
Here we re-derive neutral entropy production in a general frame in order to highlight the physical transport coefficients and their constraints. We write $T_{(1)}^{\mu\nu}$ for the first-derivative correction in
\eqref{eq:stress}. The canonical neutral entropy current is
\begin{equation}
 TS^\mu=p u^\mu-u_\nu T^{\mu\nu},\qquad
 S^\mu=su^\mu-\frac{u_\nu}{T}T_{(1)}^{\mu\nu}.
\end{equation}
The energy equation gives
$D\epsilon+w\Theta=u_\nu\nabla_\mu T_{(1)}^{\mu\nu}$.
Using $d\epsilon=T\,ds$ and $w=Ts$, one therefore obtains
\begin{align}
 \nabla_\mu(su^\mu)&=
       \frac{u_\nu}{T}\nabla_\mu T_{(1)}^{\mu\nu},\\
 \nabla_\mu S^\mu&=-T_{(1)}^{\mu\nu}
                      \nabla_\mu\left(\frac{u_\nu}{T}\right).
 \label{app:eq:neutral-entropy-divergence}
\end{align}
Using the operators of Section~\ref{sec:constitutive}, direct contraction
of \eqref{eq:stress} in \eqref{app:eq:neutral-entropy-divergence} yields
\begin{align}
 T\nabla_\mu S^\mu={}&-\varepsilon_1(D\ln T)^2
 -(\varepsilon_2+\pi_1)(D\ln T)\Theta-\pi_2\Theta^2\nonumber\\
 &-\theta_1a_\mu a^\mu-(\theta_1+\theta_2)a_\mu\perpgrad^\mu\ln T
       -\theta_2(\perpgrad\ln T)^2\nonumber\\
 &+\frac\eta2\sigma_{\mu\nu}\sigma^{\mu\nu}+O(\partial^3).
\end{align}
The ideal energy and momentum equations imply
\begin{equation}
 D\ln T=-x\Theta+O(\partial^2),\qquad
 a^\mu+\perpgrad^\mu\ln T=O(\partial^2).
\end{equation}
The vector contribution is $-\mathcal Q_\mu(a^\mu+\perpgrad^\mu\ln T)$ and is consequently
$O(\partial^3)$ even when ${\theta_1}\ne{\theta_2}$. Thus
\begin{equation}\label{app:eq:neutral-entropy-final}
 T\nabla_\mu S^\mu=\zeta_*\Theta^2
       +\frac\eta2\sigma_{\mu\nu}\sigma^{\mu\nu}+O(\partial^3),
\end{equation}
where $\zeta_*$ is the neutral bulk viscosity frame-invariant in
\eqref{eq:quartic-data}. This derives that entropy relation from the
canonical current.
The squares in \eqref{app:eq:neutral-entropy-final} are nonnegative, since the shear
tensor is transverse. Matching the prescribed $\zeta$ by
$\zeta_*=\zeta\geq0$, together with $\eta\geq0$, proves the required
second law at this order for on-shell configurations.

\subsubsection{Neutral rest-frame sound matrix and polynomial}\label{app:neutral-matrix}
To obtain the matrix of linearised perturbations and sound polynomial, we linearize around the fluid at rest using longitudinal perturbations
$(\delta T/T_0,\delta u_\parallel)$. With $q=k^2$, the conservation matrix is
\begin{equation}\label{app:eq:neutral-matrix}
 M(\varsigma,k)=\begin{pmatrix}
 {\varepsilon_1}\varsigma^2+\dfrac{w}{x}\varsigma-{\theta_2}k^2
 &ik\,[w+({\varepsilon_2}+{\theta_1})\varsigma]\\[3pt]
 ik\,[w+({\theta_2}+{\pi_1})\varsigma]
 &{\theta_1}\varsigma^2+w\varsigma+(\frac43\eta-{\pi_2})k^2
 \end{pmatrix}.
\end{equation}
Expanding $P_N(\varsigma,q)=x\det M(\varsigma,k)$ and using the neutral
bulk invariant gives exactly \eqref{eq:neutral-sound-polynomial} with
coefficients \eqref{eq:quartic-data}. The positive normalization leaves
the roots unchanged.
In particular, setting $\pi_2=4\eta/3$ makes $F_L$ in
\eqref{eq:quartic-data} vanish, but leaves the physical sound attenuation
unchanged. Expanding in small-$k$ gives the usual sound mode
\begin{equation}
 \omega_\pm=\pm\sqrt{x}\,k-i\frac{\Gamma}{2w}k^2+O(k^3).
\end{equation}
The rest frame gaps are those in Section~\ref{sec:construction-strategy}.
Other gaps must  instead be extracted after boosting, as in Section~\ref{app:sec:neutral-criterion}.

\subsubsection{Neutral rest-frame stability and causality conditions}
\label{app:sec:neutral-construction}
We now derive the reduced sound conditions used in
Section~\ref{sec:construction-strategy}. We first consider $\eta>0$, so
entropy matching gives $\Gamma=4\eta/3+\zeta>0$. The conditions
$\varepsilon_1,\theta_1>0$ follow from damping of the two rest-frame gaps
in \eqref{eq:neutral-rest-gaps}.\footnote{\label{fn:neutral-zero-transport}At
$\eta=0$, $\zeta>0$, the sound polynomial of families \family{1}-\family{3}
reduces to a stable and causal cubic polynomial. The regular families \family{4}-\family{6}
prescriptions retain a quartic, so the stability and causality proof discussed in this appendix still applies.
When both viscosities vanish, the regular families \family{1}, \family{2} and
\family{4}-\family{6} prescriptions reduce to ideal sound polynomial. Family
\family{3} can retain $\varepsilon_1>0$, which adds a damped relaxation
mode. These conclusions follow from the reduced polynomials, whose
remaining modes are linearly stable and causal in every inertial frame.}
The shear sector additionally requires
\eqref{eq:strategy-shear-bound}, as derived in
Section~\ref{sec:strategy-shear} and proved for arbitrary boosts in
Section~\ref{app:sec:shear}. We therefore focus below on the longitudinal
polynomial. The family constructions follow in
Sections~\ref{app:sec:neutral-family-derivations} and
\ref{app:sec:neutral-conformal-derivations}.

\subsubsection*{Reducing the Routh-Hurwitz inequalities for every momentum}
To reduce the inequalities, we use $R,\Sigma,\Phi,\Lh$ from \eqref{eq:invariants} and the Hurwitz
determinants \eqref{eq:rh}. Their highest-order terms follow from the identity
\begin{equation}\label{app:eq:neutral-design-RH}
 A_LB_L^2\Lh=D_L(B_LC_L-A_LD_L)-B_L^2F_L.
\end{equation}
For example, expanding
$D_Lq\{B_L(w^2+C_Lq)-A_LD_Lq\}-B_L^2(xw^2q+F_Lq^2)$
and using $D_L-xB_L=w\Gamma$ gives $\Delta_3$ in \eqref{eq:rh}.
With $a_0(q)=xw^2q+F_Lq^2$, the Routh array is
\begin{equation}
 \begin{array}{c|ccc}
 \varsigma^4&A_L&w^2+C_Lq&a_0(q)\\
 \varsigma^3&B_L&D_Lq&0\\
 \varsigma^2&\Delta_2/B_L&a_0(q)&0\\
 \varsigma^1&\Delta_3/\Delta_2&0&0\\
 \varsigma^0&a_0(q)&0&0
 \end{array}.
\end{equation}
For each $q>0$, strict stability requires the first column to be positive.
Under the stated positivity assumptions, imposing this for
\emph{every} $q>0$ gives
\begin{equation}\label{app:eq:neutral-design-unreduced}
 F_L\geq0,\qquad B_LC_L-A_LD_L\geq0,\qquad \Lh\geq0.
\end{equation}
For example, a negative coefficient of the highest power of $q$ in
$a_0$, $\Delta_2$, or $\Delta_3$ eventually violates the corresponding
condition. Conversely, the positive lower-power terms ensure strict
positivity at every finite $q>0$ even if a displayed inequality saturates.
Because $D_L>0$, the first and third conditions imply the second.
Thus the rest-frame sound stability test reduces to
$\Phi,\Lh\geq0$ in \eqref{eq:certificate}.

\subsubsection*{Adding the sound causality conditions}
The sound characteristic polynomial is $A_Ly^2-C_Ly+F_L$.
Its two roots are real and lie in $[0,1]$ if and only if
\begin{equation}\label{app:eq:neutral-design-causal}
 F_L\geq0,\quad C_L\geq0,\quad C_L^2\geq4A_LF_L,\quad
 C_L\leq2A_L,\quad A_L-C_L+F_L\geq0.
\end{equation}
Indeed, the first three conditions give real nonnegative roots.
The last two give $y_-+y_+\leq2$ and
$(1-y_-)(1-y_+)\geq0$, which force both roots to be at most one.
This argument includes the boundary roots $0$ and $1$ without replacing a
strict stability criterion by a non-strict one.
Moreover,
\begin{equation}\label{app:eq:neutral-design-interlacing}
 A_LR^2-C_LR+F_L=-A_L\Lh\leq0.
\end{equation}
With $R>x>0$ and $F_L\geq0$, this already implies reality,
nonnegativity, and $y_-\leq R\leq y_+$. Hence after solving the
Routh-Hurwitz conditions, the remaining sound speed restrictions can be
imposed as $C_L\leq2A_L$ and $A_L-C_L+F_L\geq0$.

On the branch $\pi_2=4\eta/3$, one has $\Phi=0$ and the sound conditions
reduce to $x<R\leq\Sigma\leq1$. Setting $\Sigma=1$ fixes
$\theta_2$ through \eqref{eq:luminal-design} and gives
\begin{equation}\label{app:eq:neutral-design-principal-factor}
 \begin{aligned}
 P_{N,\rm pr}&=A_L\varsigma^2(\varsigma^2+q),\\
 P_N&=P_{N,\rm pr}+B_L\varsigma(\varsigma^2+Rq)+w^2(\varsigma^2+xq).
 \end{aligned}
\end{equation}
Only the principal part factorizes in this way. The remaining terms
retain sound attenuation. Indeed, division by $\varsigma^2+xq$ leaves a
nonzero $\varsigma q$ coefficient $D_L-xB_L=w\Gamma$ from
\eqref{eq:quartic-data}, ruling out an undamped ideal sound factor when
$\Gamma>0$.

\subsubsection{Neutral families \family{1}--\family{3}}
\label{app:sec:neutral-family-derivations}
We now verify the prescriptions \eqref{eq:S-seeds} using the reduced
$\Phi=0$ conditions from Section~\ref{sec:neutral-branches-overview}:
$x<R\leq\Sigma\leq1$ and $\theta_1\geq\eta$.

\subsubsection*{Family \family{1}}
\label{app:family:S2}
The choices $\pi_2=4\eta/3$ and $\varepsilon_2=0$ make the second
column of \eqref{app:eq:neutral-matrix} divisible by
$w+\theta_1\varsigma$. Taking the determinant proves
\eqref{eq:relax-factor} for arbitrary $\theta_2$ while
entropy matching \eqref{eq:matching} gives $x\pi_1-x^2\varepsilon_1=\Gamma$.
The cubic factor in \eqref{eq:relax-factor} has only one nontrivial
Routh-Hurwitz inequality:
\begin{equation}
 w(x\pi_1q)-(x\varepsilon_1)(xwq)=w\Gamma q>0.
\end{equation}
The sound and shear speed bounds give the parameter range in
\eqref{eq:S-seeds}, with $\theta_2=0$ selecting \family{1}.
The cubic calculation proves rest-frame stability after extracting the
damped factor $w+\theta_1\varsigma$. To apply the common boosted proof in
Section~\ref{app:sec:neutral-criterion}, we also express the result in
terms of the full quartic \eqref{eq:relax-factor}. It has $\Phi=0$, and
\begin{equation}\label{app:eq:neutral-s2-speeds}
 \Sigma-R=\frac{\Gamma\theta_1}
 {x\varepsilon_1(x\varepsilon_1+\theta_1)}>0.
\end{equation}
Hence $\Lh=R(\Sigma-R)>0$. Together with the sound and shear bounds
above, this verifies \eqref{eq:certificate} for the full quartic.

\subsubsection*{Family \family{2}}
\label{app:family:S3}
Instead setting $\theta_2=\theta_1$ gives the family \family{2} in \eqref{eq:S-seeds}.
Because $\varepsilon_2=0$ and $\pi_2=4\eta/3$, neither $C_L$ nor $F_L$
depends on $\theta_2$. The entropy invariant, full sound determinant
and shear polynomial are therefore unchanged. The \family{1} proof given above applies,
including its boosted polynomial and saturated bounds.

\subsubsection*{Family \family{3}}
\label{app:family:S4}
For this family, we substitute $\theta_1=\eta$ and $\varepsilon_2=-2\eta$ into
\eqref{eq:luminal-design}. Entropy matching \eqref{eq:matching} and $\Sigma=1$ then give
the \family{3} coefficients in \eqref{eq:S-seeds}. The remaining condition
$R\leq1$ is
\begin{equation}\label{app:eq:neutral-s4-bound}
 x(1-x)\varepsilon_1\geq\Gamma-\eta(1-x)
       =\zeta+\eta(x+1/3)>0.
\end{equation}
Substitution into the original determinant gives
\begin{equation}
 {\varepsilon_2}({\theta_2}+{\pi_1})+{\theta_1}{\pi_1}
       =\eta{\varepsilon_1},
\end{equation}
so $\Sigma=1$. The inequality $R\leq1$ is precisely
\eqref{app:eq:neutral-s4-bound}. At equality $\Lh=0$, but
$\Delta_3=B_Lw^3\Gamma q>0$ for every finite $q>0$.
The boosted proof in Section~\ref{app:sec:neutral-criterion} includes this
endpoint.\footnote{\label{fn:F3-large-k-damping}At the saturated endpoint
$R=\Sigma=1$, the propagating sound modes obey
$\Re\varsigma_\pm=-B_Lw^2(1-x)/(2A_L^2k^2)+O(k^{-3})$.
Their damping therefore tends to zero as $k^{-2}$ at large momentum.}
The nonzero sound and shear speeds become equal to the speed of light.

\subsubsection*{Why \family{1}--\family{3} do not acquire a viscous conformal limit}
\label{app:sec:numbered-conformal-restriction}
The reduced inequalities do not impose a well defined conformal limit. In fact, the
coefficients just derived have the exact residuals
\begin{equation}\label{app:eq:numbered-conformal-residuals}
 \varepsilon_2-3\pi_2=
 \begin{cases}
 -4\eta,&\text{families \family{1} and \family{2}},\\
 -6\eta,&\text{\family{3}}.
 \end{cases}
\end{equation}
Therefore no choice of their free parameters can impose all conformal
coefficient relations \eqref{eq:conformal} at the conformal limit with $\eta_c>0$.
The family \family{2} obeys $\theta_1=\theta_2$, but that vector relation does not repair its
scalar mismatch. The charged uplift retains the seed coefficients
$\varepsilon_2$ and $\pi_2$, so the same residual excludes the charged families \family{1}-\family{3}
with a conformal limit.

\subsubsection{Neutral families \family{4}--\family{6} and conformal limits}
\label{app:sec:neutral-conformal-derivations}
We now verify the three constructions with finite conformal limits in
Section~\ref{sec:conformal-construction}, then discuss the overlaps between different families.\footnote{\label{fn:exact-conformal-seed}Alternatively,
we can impose the conformal relations \eqref{eq:conformal} before solving
the inequalities. At $x=1/3$ and $\zeta=0$, the neutral parametrization
in Step~3 of Section~\ref{sec:constructionstrategy} leaves $\Pi,\theta$.
For $\eta>0$, the bounds $\Pi\geq4\eta$ and
$\theta\geq3\eta\Pi/(\Pi-\eta)$ satisfy \eqref{eq:certificate}.
An exactly conformal theory also requires $\Pi,\theta,\eta\propto T^3$.
At $\eta=\zeta=0$, any $\Pi,\theta\geq0$ gives an ideal sound factor
and a stable causal quadratic, which reduces further to a damped linear
factor or a constant if $\Pi\theta=0$.}

\subsubsection*{General luminal family \family{4}}
\label{app:family:AG}
For the coefficients and parameter bounds in \eqref{eq:AG},
$\varepsilon_2=(1-f)S_0\Gamma>0$ when $\Gamma>0$, so $\theta_2$ is regular.
Substitution into \eqref{eq:quartic-data} gives $\Sigma=1$ and $\Phi=0$.
Moreover,
\begin{equation}
 x\varepsilon_1+\theta_1\geq\frac{2\Gamma}{1-x}+\theta_1,
 \qquad x<R\leq\frac{1+x}{2}<1.
\end{equation}
Hence $\Lh=R(1-R)>0$. The bound $K\geq3/4$ gives
$\theta_1\geq\eta$ because $\Gamma\geq4\eta/3$. This verifies
\eqref{eq:certificate} directly.
A well defined conformal limit \eqref{eq:conformal} further requires the parameter limits derived in
Section~\ref{sec:conformal-construction}.

\subsubsection*{Family \family{5}: strict subluminality}
\label{app:sec:charged-strict-conformal}
We check the strict limits for the deformation of the \family{5} family
constructed in Section~\ref{sec:conformal-construction}. Substituting \eqref{eq:balanced} and the \family{5} vector coefficients
into \eqref{eq:quartic-data} gives $\Sigma=\rho$ and $\Phi=0$, with
$\rho_0,\rho$ defined in Section~\ref{sec:conformal-construction}.
For $\Gamma>0$, the distance to the causal bound is
\begin{align}\label{app:eq:charged-AS-sound-margin}
 \rho-\rho_0&=\frac{m(\theta_2-\theta_1)}{\varepsilon_1\theta_1}
 =\frac{\chi(\lambda-1)(1-x)}{2\lambda},\nonumber\\
 1-\rho&=\frac{(\lambda-1)(1-x)(1-\chi)}{2\lambda}>0,\\
 1-\chi&=\frac{1+(1-\beta)(3x-1)^2}{1+(3x-1)^2}>0.\nonumber
\end{align}
To prove the remaining stability inequality, we introduce the positive denominator
\begin{equation}
 d_5=8x(1-x)+\lambda(1+x)^2>0,\qquad
 R=\frac{D_L}{B_L}=x+\frac{4x(1-x)^2}{d_5}.
\end{equation}
Direct subtraction gives the strictly positive expression
\begin{equation}\label{app:eq:charged-AS-R-margin}
 \rho_0-R=
 \frac{(1-x)\big[8x(1-x)+\lambda(\lambda+1)(1+x)^2\big]}
      {2\lambda d_5}>0.
\end{equation}
Hence
\begin{equation}\label{app:eq:charged-AS-strict-interval}
 x<R<\rho_0\leq\rho<1,\qquad
 \Lh=R(\rho-R)>0.
\end{equation}
The shear speed also has a strict bound. Since $\Gamma\geq4\eta/3$,
\begin{equation}\label{app:eq:charged-AS-shear-margin}
 0\leq y_T=\frac{\eta}{\theta_1}
 \leq\frac{3x(1-x)^2}{\lambda(1+x)^2}<\frac1\lambda<1.
\end{equation}
The strict middle inequality follows from
$(1+x)^2-3x(1-x)^2=(1-x)+x^2(7-3x)>0$.
The sound and shear characteristic speeds therefore have magnitude
strictly less than one while one of the sound branches has vanishing characteristic speed.
Entropy matching \eqref{eq:matching} follows from
$-\pi_2-x^2\varepsilon_1+2xm=\Gamma-4\eta/3=\zeta$. We have therefore verified \eqref{eq:certificate} for this family. The conformal limits of this family are given in Section~\ref{sec:conformal-construction}.
The remaining residual is $\theta_2-\theta_1=\chi(\lambda-1)m$.
For $\eta_c>0$ and finite $\lambda_c>1$, it vanishes only if
$\chi\to0$, as ensured by \eqref{eq:chi}. At $\lambda_c=1$ the upper
sound speed becomes luminal.

\subsubsection*{Family \family{6}: nonzero characteristic speeds}
\label{app:sec:charged-nonzero-conformal}
Here we prove the bounds \eqref{eq:AN-bounds} for the neutral coefficients
\eqref{eq:AN}, with $\eta>0$ and $\zeta\geq0$. The coefficients
$\varepsilon_1,\varepsilon_2,\pi_1$ in \eqref{eq:AN} are obtained by
multiplying their values in \eqref{eq:balanced} by $\Lambda$, with
$\varepsilon_2=\pi_1$ throughout. Consequently,
$-x^2\varepsilon_1+2x\varepsilon_2=\Lambda\Gamma$.
Entropy matching \eqref{eq:matching} fixes $\pi_2$ through
$-\pi_2+\Lambda\Gamma=\zeta$, preserving the prescribed bulk viscosity.
It also gives $F_L=x\theta_2(\Lambda-1)\Gamma>0$.
We now check that the vector coefficients leave sufficient margins for
both stability and causality. We write
$\Phi_0=(\Lambda-1)x(1-x)/(2\Lambda)$ for the value at $\chi=0$,
with $\Sigma_0$ as defined in \eqref{eq:AN}. Substitution into the original neutral matrix
\eqref{app:eq:neutral-matrix} gives the invariants \eqref{eq:AN-speeds}.
For the remaining inequality we set
\begin{equation}
 M=1+6x-3x^2>0,\qquad
 j=\frac{2(\Lambda-1)x(1-x)}{\Lambda(1+x)}.
\end{equation}
Thus $\Phi=\Phi_0+j\chi(1-\Sigma_0)$.
We define
\begin{align}\label{app:eq:charged-AN-positive-Q}
 Q={}&(\Lambda-1)^2(3x-1)^2M^2\nonumber\\
 &+2(\Lambda-1)(1+x)M[5(1-x)+x^2(19-3x)]\nonumber\\
 &+(3-x)(1+x)^3(3+10x-x^2),\\
 \Lh_0={}&\frac{x(1-x)Q}{4\Lambda^2(1+x)^2M^2}>0.\nonumber
\end{align}
Every bracket is positive on $0<x<1$, and the last term in $Q$
is strictly positive. An algebra calculation gives
\begin{equation}\label{app:eq:charged-AN-positive-L}
 R(\Sigma_0-R)-\Phi_0=\Lh_0,\qquad
 \Lh
   =\Lh_0+\chi(1-\Sigma_0)(R-j)>0.
\end{equation}
The final sign follows from the explicit bounds
\begin{align}
 1-\Sigma_0&=\frac{1-x}{4}
     +\frac{2(\Lambda-1)x(1-x)^2}{\Lambda(1+x)^2}>0,\\
 x-j&=\frac{x[2-\Lambda+(3\Lambda-2)x]}{\Lambda(1+x)}>0,
 \qquad R>x,\nonumber\\
 1-\Sigma&=(1-\chi)(1-\Sigma_0)>0,\qquad \Phi>0.\nonumber
\end{align}
The upper restriction $\Lambda\leq2$ is a sufficient bound ensuring
$x-j>0$ for all $x$, including the non-thermodynamic case.

Now define $f(y)=y^2-\Sigma y+\Phi$. Then $f(0)=\Phi>0$,
$f(R)=-\Lh<0$, and $R>0$. Hence $f$ has two distinct positive roots
$y_-<R<y_+$. Their sum is $\Sigma<1$, which proves the sound bounds in \eqref{eq:AN-bounds}.
In particular $\Sigma-R>0$. The shear squared speed obeys
\begin{equation}
 y_T=\frac{\eta}{\theta_1}
 \leq\frac{3x(1-x)^2}{2\Lambda(1+x)^2}
 <\frac1{2\Lambda}<1.
\end{equation}
The strict upper bound uses the same positive polynomial as
\eqref{app:eq:charged-AS-shear-margin}. With $\eta>0$, all sound and shear
characteristic speeds are nonzero. These inequalities verify
\eqref{eq:certificate}.

At a regular conformal endpoint, multiplying these three coefficients
by a finite $\Lambda_c$ gives $\Pi=4\Lambda_c\eta_c$. The value of
$\pi_2$ fixed by entropy matching also tends to $\Pi/3$, giving the
conformal scalar relations in Section~\ref{sec:conformal-construction}.
Conformality does not select $\Lambda=2$: the range $1<\Lambda\leq2$
comes from the stability bounds.
The vector residual has endpoint
$\theta_{2,c}-\theta_{1,c}=4\eta_c(2\Lambda_c-1)\chi_c$.
For $\eta_c>0$ and $\Lambda_c>1$, conformal symmetry therefore requires
$\chi_c=0$. A limit
$\Lambda_c=1$ reaches \family{5} with $\lambda=2$ and loses the nonzero lower
sound speed.

\subsubsection*{Overlaps between the families \family{4}, \family{5} and \family{6}}
\label{app:sec:family-inclusions}
Substituting the parameters \eqref{eq:AG-equal-vector} into the family
\family{4} prescription \eqref{eq:AG} gives the values of
$\varepsilon_1$, $\varepsilon_2=\pi_1=m$ and $\pi_2$ in
\eqref{eq:balanced}, together with $\theta_1=\theta_2=\theta_0$.
They lie in the \family{4} domain: the scalar lower bound is saturated,
$f=1/2$, and $K>3/4$ follows from the polynomial positivity used in
\eqref{app:eq:charged-AS-shear-margin}. Setting $\lambda=1$ in \family{5} gives this member of \family{4},
independently of $\beta$, proving the first overlap stated
in Section~\ref{sec:catalogue}.
For \family{6}, setting $\Lambda=1$ gives the same scalar coefficients
\eqref{eq:balanced} and
\[
 \pi_2=4\eta/3,\qquad \Sigma_0=(3+x)/4,\qquad
 \theta_1=2\theta_0,\qquad \theta_2=2\theta_0+\chi m,
\]
which is exactly \family{5} at $\lambda=2$ and the same $\beta$.
At $\Gamma>0$, matching \family{5}'s value $\pi_2=4\eta/3$ to \family{6} forces
$\Lambda=1$. Matching $\theta_1$ then forces $\lambda=2$.
Thus this overlap with \family{5} occurs only at $\Lambda=1$ in \family{6}, and
only at $\lambda=2$.

\subsection{Charged frames}\label{app:charged-frames}
We now construct charged frames from the neutral ones. We first recall
the charged thermodynamic conditions, then derive the coefficient map
and check its entropy matching, factorization and remaining freedom.

\subsubsection{Thermodynamic domain and susceptibilities}\label{app:assumptions}
We use the full domain \eqref{eq:domain}. The charged uplift evaluates
the neutral seed at the charged values in $\zeta_*$ in
\eqref{eq:effective}. We now derive the fluctuation-positivity inequalities in
\eqref{eq:domain} from equilibrium thermodynamics.
Let $\delta_p=p_{TT}p_{\mu\mu}-p_{T\mu}^2$ denote the pressure-Hessian
determinant, with entropy density $s=p_T$ as in
Section~\ref{sec:constitutive}. Positivity of equilibrium
energy-charge fluctuations makes the pressure Hessian positive
semidefinite, equivalently $n_\mu\geq0$, $E\geq0$, $\Delta_J\geq0$,
as in \cite{HoultKovtun}. On our regular domain
it is positive definite and direct differentiation gives
\[
 \begin{gathered}
 n_\mu=p_{\mu\mu}>0,\qquad \Delta_J=T^2\delta_p>0,\\
 E=T(p_{TT}+2\alpha p_{T\mu}+\alpha^2p_{\mu\mu})>0,\qquad
 wx=\frac{p_{\mu\mu}s^2-2p_{T\mu}sn+p_{TT}n^2}{\delta_p}.
 \end{gathered}
\]
Consequently
\[
 wx n_\mu-n^2=\frac{(p_{\mu\mu}s-p_{T\mu}n)^2}{\delta_p}\geq0,
\]
which proves the last inequality in \eqref{eq:domain}. Using $x=A+rTB$ from \eqref{eq:thermo}, the other inequality
follows from
\[
 A+\frac{n_\mu}{w}(TB)^2-rTB
 =\frac{n_\mu}{w}\left(TB-\frac{n}{n_\mu}\right)^2
   +x-\frac{n^2}{wn_\mu}\geq0.
\]
The condition $\Delta_J>0$ selects regular susceptibilities, and $0<x<1$
restricts the ideal sound cone. We do not consider the case of singular susceptibilities nor vanishing/light speeds
$x=0,1$.

\subsubsection{Charged entropy production and the coefficient uplift}
\label{app:sec:charged-uplift}
We now derive the map \eqref{eq:uplift-stress}-\eqref{eq:uplift-current} for an
arbitrary neutral seed satisfying the preceding stability and causality requirements, and prove
\eqref{eq:factorization} directly using the original charged matrix. We use the thermodynamic domain
\eqref{eq:domain}, transport constraints \eqref{eq:thermo-domain} and
susceptibilities \eqref{eq:suscept}.

\subsubsection*{Charged entropy production and matching identities}
\label{app:sec:charged-entropy}
We begin by re-deriving the entropy constraints in an arbitrary frame stated in Section~\ref{sec:overview}. For the canonical current
$S^\mu=(p u^\mu-u_\nu T^{\mu\nu}-\mu J^\mu)/T$, conservation and
thermodynamics give
\begin{equation}
 \nabla_\mu S^\mu=-T_{(1)}^{\mu\nu}\nabla_\mu(u_\nu/T)
                         -J_{(1)}^\mu\nabla_\mu\alpha,
\end{equation}
where $J_{(1)}^\mu$ is the first-derivative correction in \eqref{eq:current}.
Using the scalar and vector corrections of \eqref{eq:scalars}--\eqref{eq:vectors},
the contraction is
\begin{equation}\label{app:charged-entropy-contraction}
 \begin{aligned}
 T\nabla_\mu S^\mu={}&-\mathcal E D\ln T-\mathcal P\Theta-T\mathcal N D\alpha\\
 &-\mathcal Q_\mu(a^\mu+\perpgrad^\mu\ln T)
 -T\mathcal J_\mu\perpgrad^\mu\alpha
 +\frac\eta2\sigma_{\mu\nu}\sigma^{\mu\nu}+O(\partial^3).
 \end{aligned}
\end{equation}
To reduce the entropy production using the ideal conservation equations,
we first express $A$ and $B$ in terms of the susceptibilities
\eqref{eq:suscept}. The pressure identities
$(\partial_Tp)_\alpha=w/T$ and $(\partial_\alpha p)_T=Tn$, together
with the Maxwell relation $F=T^2N$, give
\begin{equation}\label{app:eq:charged-uplift-thermo}
 A=\frac{wq_\alpha}{T{\Delta_J}},\qquad
 B=-\frac{wq_T}{{\Delta_J}}.
\end{equation}
These follow by expressing the derivatives in \eqref{eq:thermo} as
Jacobians with respect to $(T,\alpha)$, whose transformation to
$(\epsilon,n)$ has determinant $\Delta_J$.
The ideal scalar equations
$E D T+F D\alpha=-w\Theta$ and $N D T+H D\alpha=-n\Theta$
give $D\ln T=-A\Theta$ and $D\alpha=-B\Theta$ on inversion.
Ideal momentum conservation gives
$a^\mu=-\perpgrad^\mu\ln T-rT\perpgrad^\mu\alpha$.

Substituting the constitutive relations gives
the following quadratic expression, before imposing any frame-coefficient
relation to cancel a contribution:
\begin{equation}\label{eq:entropy}
 \begin{aligned}
 \nabla_\mu S^\mu
 ={}&\frac{\eta}{2T}\sigma_{\mu\nu}\sigma^{\mu\nu}
 -\frac1T\Bigl[\eps_1(D\ln T)^2+(\eps_2+\pi_1)(D\ln T)\Theta+\pi_2\Theta^2\\
 &\quad +(\eps_3+T\nu_1)(D\ln T)(D\alpha)
          +(\pi_3+T\nu_2)\Theta D\alpha+T\nu_3(D\alpha)^2\\
 &\quad +\theta_1a_\mu a^\mu
          +(\theta_1+\theta_2)a_\mu\perpgrad^\mu\ln T
          +\theta_2(\perpgrad\ln T)^2\\
 &\quad +(\theta_3+T\gamma_1)a_\mu\perpgrad^\mu\alpha
          +(\theta_3+T\gamma_2)\perpgrad_\mu\ln T\,\perpgrad^\mu\alpha\\
 &\quad +T\gamma_3(\perpgrad\alpha)^2\Bigr]
          +\OO(\partial^3)\\
 ={}&\frac{\eta}{2T}\sigma_{\mu\nu}\sigma^{\mu\nu}
       +\frac\zeta T\Theta^2+\cD T(\perpgrad\alpha)^2\\
 &\quad +\bigl[r(\theta_2-\theta_1)-(\gamma_2-\gamma_1)\bigr]
              \perpgrad_\mu\ln T\,\perpgrad^\mu\alpha+\OO(\partial^3).
 \end{aligned}
\end{equation}
Only the last equality in \eqref{eq:entropy} uses these
ideal equation relations, together with the definitions in
\eqref{eq:matching}. The mixed term
has not been set to zero. There is no remaining
$(\perpgrad\ln T)^2$ term, so positivity for arbitrary independent
$\perpgrad^\mu\ln T$ and $\perpgrad^\mu\alpha$ requires the vector
relation in \eqref{eq:thermo-domain}. Imposing that equation makes
the displayed quadratic production nonnegative. This condition permits unequal
$\theta_1,\theta_2$ and unequal $\gamma_1,\gamma_2$. It does not impose
a thermodynamic frame.

\subsubsection*{Derivation of the coefficient map and entropy matching}
\label{app:sec:charged-map-entropy}
We start with a set of neutral coefficients or one the six neutral families we found but  evaluated at the
charged values in  \eqref{eq:effective}. We replace its temperature
gradient by the pressure gradient using
\begin{equation}
 \frac{dp}{w}=\frac{dT}{T}+rT\,d\alpha.
\end{equation}
This replacement yields the third stress coefficients in
\eqref{eq:uplift-stress}. We then choose the current coefficients so that,
after linearising about uniform equilibrium and Fourier transforming in its
rest frame, the fluctuation $h=\delta n-r\delta\epsilon$ obeys the independent
telegrapher equation
$F_q(\varsigma,q)h=(\tau_q\varsigma^2+\varsigma+D_qq)h=0$, as described
in Step~4 of Section~\ref{sec:constructionstrategy}, with $F_q$ defined in
\eqref{eq:strategy-factorization}. We obtain this equation by subtracting
$r$ times the linearised energy-conservation equation from the
charge-conservation equation, with $r=n/w$ evaluated at equilibrium.
Imposing this condition together with entropy matching \eqref{eq:matching}
gives the current coefficients \eqref{eq:uplift-current}. We verify the
corresponding matrix identity \eqref{app:eq:charged-uplift-row} in the
following subsection.

To check entropy matching \eqref{eq:matching}, we first combine
\eqref{app:eq:charged-uplift-thermo} with $x=A+rTB$ to obtain
$x=w(q_\alpha-rT^2q_T)/(T\Delta_J)=w\Xi/(T\Delta_J)$.
This proves the susceptibility identity in \eqref{eq:effective}, with
$\Xi>0$ on the domain \eqref{eq:domain}, and gives
$D_q=\mathcal D w/(\Delta_Jx)$.
The vector matching relations in \eqref{eq:matching} then follow
directly from the map:
\begin{align}
 \gamma_2-\gamma_1&=r(\theta_2-\theta_1),\\
 r(\theta_3-rT\theta_1)+rT\gamma_1-\gamma_3
     &=D_q\Xi=\mathcal D T.
\end{align}
Substituting the map into the charged bulk invariant, using
$q_\alpha B+Tq_TA=0$ from \eqref{app:eq:charged-uplift-thermo} and
$A+rTB=x$, gives
\begin{equation}\label{app:eq:charged-uplift-bulk}
 \zeta=-\pi_2-x^2\varepsilon_1+x(\varepsilon_2+\pi_1)
       -\frac{D_qT^2\Delta_JB^2}{w}
 =\zeta_*-\frac{\mathcal D T^2B^2}{x}.
\end{equation}
This derives the charge values  in \eqref{eq:effective} from
\eqref{eq:matching}.  These identities cancel the mixed term in \eqref{eq:entropy} and leave
the three nonnegative terms with the prescribed $\eta,\zeta,\mathcal D$.

\subsubsection*{Exact factorization of the original charged matrix}
The purpose of this subsection is to verify the map introduced above directly from the linearised matrix. With equation order energy, longitudinal momentum, charge and
unknowns $(\delta T,\delta\alpha,\delta u_\parallel)$, the original sound matrix is
\begingroup\small
\begin{equation}\label{app:original-charged-matrix}
M_{\rm charged}(\varsigma,k)=
\begin{pmatrix}
 E\varsigma+\varepsilon_1\varsigma^2/T-\theta_2q/T&
 F\varsigma+\varepsilon_3\varsigma^2-\theta_3q&ik[w+(\varepsilon_2+\theta_1)\varsigma]\\
 ik[w/T+(\theta_2+\pi_1)\varsigma/T]&
 ik[Tn+(\theta_3+\pi_3)\varsigma]&\theta_1\varsigma^2+w\varsigma+(4\eta/3-\pi_2)q\\
 N\varsigma+\nu_1\varsigma^2/T-\gamma_2q/T&
 H\varsigma+\nu_3\varsigma^2-\gamma_3q&ik[n+(\nu_2+\gamma_1)\varsigma]
\end{pmatrix}.
\end{equation}
\endgroup
Here $E,F,N,H$ are the derivatives at fixed $\alpha$ or $T$ defined
in \eqref{eq:suscept}. Substituting $\omega=i\varsigma$ into the Fourier space
conservation equations gives exactly this matrix, including the minus
signs on its spatial-gradient entries.

We define $\mathcal R_\epsilon$ and $\mathcal R_n$ as the first and third rows of $M_{\rm charged}$ in \eqref{app:original-charged-matrix}, corresponding respectively to the linearised energy- and charge-conservation equations. Equations~\eqref{eq:uplift-stress}-\eqref{eq:uplift-current}
give the exact row identity
\begin{equation}\label{app:eq:charged-uplift-row}
 \mathcal R_n-r\mathcal R_\epsilon
 =F_q(\varsigma,q)\,(q_T,\ q_\alpha,\ 0).
\end{equation}
Here $F_q$ is defined in \eqref{eq:strategy-factorization}, and $\mathcal R_\epsilon$ is the row of $\partial_\mu T^{\mu 0}=0$.
Its covariant energy projection is
$u_\nu\nabla_\mu T^{\mu\nu}=-\partial_\mu T^{\mu0}$ in the rest frame.
Thus the row difference is the Fourier transform of
$\delta[\nabla_\mu J^\mu+r u_\nu\nabla_\mu T^{\mu\nu}]$ at uniform
equilibrium. It is a linear identity. 
We use $(\delta\epsilon,h)$, with $h$ from Step~4 of
Section~\ref{sec:constructionstrategy} and
$\delta\epsilon=E\delta T+F\delta\alpha$ from \eqref{eq:suscept}.
The Jacobian of this change is $E q_\alpha-Fq_T={\Delta_J}$.
On the $h=0$ subspace,
\begin{equation}
 \frac{\delta T}{T}+rT\delta\alpha=\frac{x}{w}\delta\epsilon.
\end{equation}
Consequently the remaining energy-momentum block is
\begin{equation}
 \begin{pmatrix}
 \varsigma+\dfrac{x \varepsilon_1}{w}\varsigma^2-\dfrac{x \theta_2}{w}q
 &ik[w+(\varepsilon_2+\theta_1)\varsigma]\\[3pt]
 ik\left[x+\dfrac{x(\theta_2+\pi_1)}w \varsigma\right]
 &\theta_1\varsigma^2+w\varsigma+(\frac43\eta-\pi_2)q
 \end{pmatrix}.
\end{equation}
Its determinant is $P_N/w$, where $P_N$ is exactly the neutral polynomial
\eqref{eq:neutral-sound-polynomial} with the indicated coefficient choices.
Expanding the charge row, including the row/column sign and Jacobian,
therefore proves \eqref{eq:factorization} on the original determinant.
The shear polynomial remains \eqref{eq:strategy-shear-polynomial}.

\subsubsection{Charged families and inherited properties}
\label{app:sec:family-uplift-matching}
\label{app:sec:charged-explicit}\label{app:sec:charged-limits-scope}
We now check the conformal limits and family overlaps of the charged
uplifts defined in Section~\ref{sec:charged-factorization-overview}.
For the conformal limits of \family{4}-\family{6}, the additional
relations \eqref{eq:conformal} are controlled by the differences
\begin{align}\label{app:eq:uplift-conformal-residuals}
 \varepsilon_3-3\pi_3&=rT(\varepsilon_1-3\pi_1),\nonumber\\
 \nu_1-3\nu_2&=r(\varepsilon_1-3\varepsilon_2)+T(\tau_q-3D_q)q_T,\\
 \gamma_1-\gamma_2&=r(\theta_1-\theta_2).\nonumber
\end{align}
At a regular conformal endpoint, we take the conductivity $\mathcal D$
and the parameter $\alpha_q$ to have finite limits. The coefficients
$D_q,\tau_q$ in \eqref{eq:effective} are then finite, while
$B,q_T\to0$ and $\zeta_*-\zeta\to0$.
The uplift leaves the six neutral stress coefficients unchanged.
If these have finite limits satisfying \eqref{eq:conformal}, the
differences in \eqref{app:eq:uplift-conformal-residuals} also vanish,
and all additional coefficients in
\eqref{eq:uplift-stress}--\eqref{eq:uplift-current} remain finite.
Thus the charged families \family{4}-\family{6} have finite conformal
limits under the parameter conditions given in
Section~\ref{sec:conformal-construction}. 

To evaluate the characteristic speeds at these endpoints, we substitute
the neutral limits into the quartic and retain the charge factor
\eqref{eq:strategy-factorization}. For \family{5}, $y_-=0$.
At the conformal endpoint with $\eta,\mathcal D>0$, the nonzero squared
characteristic speeds are
\begin{equation}\label{app:eq:charged-AS-conformal-speeds}
 \rho_c=\frac{2\lambda+1}{3\lambda}<1,\qquad
 y_{T,c}=\frac1{4\lambda}<1,\qquad y_{q,c}=\frac1{\alpha_q}<1.
\end{equation}
For example, $\lambda=\alpha_q=2$ gives the nonzero conformal squared
speeds $(5/6,1/8,1/2)$. The choice $\beta=0$ gives thermodynamic frames,
while $\beta=1$ gives non-thermodynamic frames for $x\ne1/3$.
Both approach the same conformal coefficients.

For \family{6}, the charged principal branches consist of two sound
pairs, one charge pair and two shear pairs. At the conformal endpoint,
with positive limiting $\eta,\mathcal D$, their squared comoving
characteristic speeds are nonzero and strictly subluminal:
\begin{equation}
 y_\pm=\frac{4\Lambda+1\pm\sqrt{24\Lambda+1}}{12\Lambda},\qquad
 y_T=\frac1{8\Lambda},\qquad y_q=\frac1{\alpha_q}.
\end{equation}
The choice $\Lambda=\alpha_q=2$ gives the result
\eqref{eq:benchmark}.

The neutral family overlaps proved in
Section~\ref{app:sec:neutral-conformal-derivations} also hold after
applying the common charged uplift at the same physical inputs and
the same $\alpha_q>1$.

\subsubsection{Additional frame freedom preserving the dispersion relations}
\label{app:sec:family-hierarchy}
The charged uplift is not unique. We now derive the additional coefficient
freedom in \eqref{eq:free-shifts} that preserves its dispersion relations
and physical transport coefficients. We retain the same
equation $F_qh=0$ for the charge fluctuation $h=\delta n-r\delta\epsilon$,
implemented by \eqref{app:eq:charged-uplift-row}. Comparing coefficients
in the original matrix gives
\begin{align}\label{app:eq:hierarchy-five}
 \nu_1-r\varepsilon_1&=T\tau_qq_T,&
 \nu_3-r\varepsilon_3&=\tau_qq_\alpha,\\
 \gamma_2-r\theta_2&=-TD_qq_T,&
 \gamma_3-r\theta_3&=-D_qq_\alpha,\nonumber\\
 \nu_2+\gamma_1&=r(\varepsilon_2+\theta_1).&&\nonumber
\end{align}
Together with the vector entropy relation in \eqref{eq:thermo-domain},
these conditions determine the current coefficients. Four retain the
form in \eqref{eq:uplift-current}, while
$\nu_3=r\varepsilon_3+\tau_qq_\alpha$ and
$\gamma_3=r\theta_3-D_qq_\alpha$ now allow $\varepsilon_3,\theta_3$
to differ from \eqref{eq:uplift-stress}.
Conductivity matching gives $\mathcal D T=D_q\Xi$, as in
\eqref{eq:effective}. We keep $\tau_q$ independent to include
zero-diffusion cases.\footnote{\label{fn:independent-charge-relaxation}At
$D_q=0$, we may retain $\tau_q>0$. Then
$F_q=\varsigma(1+\tau_q\varsigma)$ has an undamped advected mode and
a damped relaxation mode, both with zero comoving characteristic speed.
The choice $\tau_q=\alpha_qD_q$ with finite $\alpha_q$ instead sets
$\tau_q=0$. The determinant identities remain valid for the reduced
polynomials. The three-function shifts cannot remove the zero shear
speed at $\eta=0$.}
For $D_q>0$, we use the stable and causal range
in \eqref{eq:effective}.

The existing frame freedom can be parametrised by three functions. Each pair $(h_1,h_3)=(\varepsilon_1,\varepsilon_3)$,
$(\pi_1,\pi_3)$ or $(\theta_2,\theta_3)$ can be separated into a
representative obeying \eqref{eq:uplift-stress} and an additional term.
We define
\begin{equation}\label{app:eq:hierarchy-inverse}
 \bar h=\frac{q_\alpha h_1-Tq_Th_3}{\Xi},\qquad
 f_h=\frac{h_3-rTh_1}{\Xi}.
\end{equation}
Then
\begin{equation}\label{app:eq:hierarchy-decomposition}
 (h_1,h_3)=(\bar h,rT\bar h)+f_h(Tq_T,q_\alpha).
\end{equation}
The two directions $(1,rT)$ and $(Tq_T,q_\alpha)$ have determinant
$\Xi>0$, so this decomposition is unique and remains regular when
$B$ or $q_T$ vanishes.
We denote the representative coefficients by
$\bar\varepsilon_1,\bar\pi_1,\bar\theta_2$ and the three functions by
$f_E,f_P,f_Q$. The other representative coefficients are unchanged:
$\bar\varepsilon_2=\varepsilon_2$, $\bar\pi_2=\pi_2$ and
$\bar\theta_1=\theta_1$. Together with \eqref{app:eq:hierarchy-five} and the vector entropy
relation, this decomposition accounts for every choice of frame in the class
specified above. The additional terms are precisely the shifts
\eqref{eq:free-shifts}, including their associated current shifts.

\subsubsection*{Preservation of dispersion relations and physical transport}
The preceding decomposition identifies the remaining coefficient
freedom. Here we show that varying $f_E,f_P,f_Q$ at fixed representative
coefficients, thermodynamic state and charge factor $F_q$ preserves
the dispersion relations and the prescribed $\eta,\zeta,\mathcal D$.
The shifts therefore give additional frame choices with the same
frequency roots and characteristic speeds, while changing the
individual BDNK coefficients. To see why the dispersion relations are
unchanged, we use the variables $(\delta\epsilon,\delta u_\parallel,h)$
and the matrix reduction of Section~\ref{app:sec:charged-uplift}. Its last equation is $F_qh=0$,
so the last row is $(0,0,F_q)$. The shifts \eqref{eq:free-shifts} add
\begin{equation}
 (f_E\varsigma^2-f_Qq)h,\qquad
 ik\varsigma(f_P+f_Q)h,
\end{equation}
to the energy and longitudinal momentum equations, respectively.
Both additions multiply $h$, so they change only the last column of
these two equations. The last row and the energy-momentum block
acting on $(\delta\epsilon,\delta u_\parallel)$ are unchanged.
Expanding the determinant along the last row gives $F_q$ times the
determinant of this unchanged block. Restoring the normalization of the
original charged matrix therefore gives
\begin{equation}\label{app:eq:hierarchy-isospectral}
 P_{\rm new}(\varsigma,q)=P_{\rm old}(\varsigma,q)
   =-\frac{\Delta_J} w F_q(\varsigma,q)\,\bar P_N(\varsigma,q),
\end{equation}
where $\bar P_N$ is \eqref{eq:neutral-sound-polynomial} evaluated at the
representative coefficients. This proves that the longitudinal roots,
damping rates and gaps are unchanged. The shear polynomial is unchanged
because $\theta_1$ and $\eta$ are fixed. The characteristic speeds obtained
from these polynomials are consequently unchanged as well. The identity also holds for
complex $\varsigma,q$, giving the boosted statement in
footnote~\ref{fn:deformation-boosts}.

Preservation of physical transport follows from the
entropy matching identities \eqref{eq:matching}. Using
$TAq_T+Bq_\alpha=0$ from \eqref{app:eq:charged-uplift-thermo}
and $A+rTB=x$, the change in the bulk viscosity is
\begin{equation}
 \delta\zeta
 =(TAq_T+Bq_\alpha)(f_P-xf_E)=0.
\end{equation}
The conductivity changes by
$(r\delta\theta_3-\delta\gamma_3)/T=0$, and the vector entropy
relation changes by $\delta\gamma_2-r\delta\theta_2=0$.
Together with $\delta\eta=0$, these identities prove preservation of
all physical transport coefficients and entropy matching.

\subsubsection*{Preserving finite conformal limits}
We now determine which shifts preserve the finite conformal limit of a
frame that already satisfies the conformal relations.
At a regular conformal endpoint, $q_T\to0$ and
$q_\alpha\to T\Delta_J/(3w)>0$. If $f_E,f_P,f_Q$ have finite limits,
the changes in $\pi_1-3\pi_2$, $\nu_1-3\nu_2$,
$\theta_1-\theta_2$ and $\gamma_1-\gamma_2$ vanish because they contain
$q_T$. The remaining condition is $f_E-3f_P\to0$, as stated in
Section~\ref{sec:catalogue}, since
$\delta(\varepsilon_3-3\pi_3)=q_\alpha(f_E-3f_P)$.
Thus $f_Q$ may have any finite limit, while the limiting values of
$f_E$ and $f_P$ must satisfy this relation. If we also want to retain the original frame's limiting coefficients
themselves, we take all three functions to vanish at conformality.
For example, we can choose
\begin{equation}\label{app:eq:hierarchy-explicit-deformation}
 f_a=c_a\frac{\Gamma}{\Xi}
       \frac{(3x-1)^2}{1+(3x-1)^2},\qquad
 a\in\{E,P,Q\},\qquad c_a\in\mathbb R,
\end{equation}
with finite dimensionless constants $c_a$ and $\Gamma$ evaluated using
the bulk viscosity in \eqref{eq:effective}. On the regular
domain, these functions are finite and vanish as $x\to1/3$, including
at $\Gamma=0$.

\subsection{Generalisations to arbitrary boosts}\label{app:arbitrary-boosts}
The connection between causality and stability in different inertial
frames is established in Refs.~\cite{GavassinoBoosts,WangPu,BDN2022}.
Here we verify stability and causality directly for our families by
analysing the sound, shear and charge polynomials in an arbitrarily
boosted frame. We check stability for all real wavevectors, including
$k=0$, and causal propagation in every spatial direction. The proof
also covers cases in which characteristic speeds equal the speed of
light.

\subsubsection{Boosted sound quartic: gaps, stability and causality}
\label{app:sec:neutral-criterion}\label{sec:boosts}
As explained in Section~\ref{sec:construction-strategy}, boosting a
damped mode with a real wavevector in the fluid rest frame generally
makes its transformed wavevector complex. To test stability in a moving
frame, we instead prescribe a real wavevector $\boldsymbol k$ in that
frame and solve for its complex growth rate $\varsigma$. Throughout this
section, $k=|\boldsymbol k|$, and primes denote quantities in the fluid
rest frame. We first check the gaps at $k=0$. For $k>0$, we then show that a root
cannot acquire a positive real part as the boost varies: it can neither
cross the imaginary axis nor enter from infinity. Finally, we check the
characteristic speeds for every propagation angle. The common shear and charge
proof follows in Section~\ref{app:sec:shear}, and the full charged system
is treated in Section~\ref{app:sec:charged-boost-proof}.

The characteristic roots are $y_\pm$ in \eqref{eq:speeds}, so
\begin{equation}\label{app:eq:neutral-principal}
 A_Ly^2-C_Ly+F_L=A_L(y-y_-)(y-y_+).
\end{equation}
\begin{proposition}\label{app:prop:neutral}
Suppose $w>0$, $0<x<1$, $\Gamma>0$, $A_L>0$, $B_L>0$, and the
sound inequalities in \eqref{eq:certificate} hold. These are its first
three conditions, involving $\Phi$, $\Lh$ and $y_\pm$.
Then every sound root at a real wavevector with $k>0$ satisfies $\Re \varsigma<0$
for every $|v|<1$ and every propagation angle. At $k=0$ there are
two roots at $\varsigma=0$ and two gapped modes with negative real parts. All sound
characteristic velocities are real and have magnitude at most one.
\end{proposition}

Rest-frame stability follows from the earlier Routh--Hurwitz analysis
in Section~\ref{app:sec:neutral-construction} and the gaps in
Eq.~\eqref{eq:neutral-rest-gaps}. The assumptions $A_L,B_L>0$ imply
$\varepsilon_1,\theta_1>0$ by Eq.~\eqref{eq:quartic-data}. We now
establish stability under arbitrary boosts, first checking the $k=0$
gaps and then treating $k>0$.

\subsubsection*{Boosted $k=0$ gaps}
We align the boost with the $z$ axis. In the convention
$\varsigma=-i\omega$, the Lorentz transformation gives
\begin{equation}\label{app:eq:neutral-boost}
 \varsigma'=\gamma(\varsigma+ivk_z),\qquad
 q'=k_\perp^2+\gamma^2(k_z-iv\varsigma)^2,\qquad
 \gamma=(1-v^2)^{-1/2}.
\end{equation}
The boosted polynomial is $P_{N,\mathrm{boost}}(\varsigma,k_z,k_\perp)=P_N(\varsigma',q')$.
Here $k_\perp$ is the magnitude of the wavevector perpendicular to the
boost. At $k=0$, we have $\varsigma'=\gamma\varsigma$ and
$q'=-\gamma^2v^2\varsigma^2$. Substituting these into
\eqref{eq:neutral-sound-polynomial} and removing the factor
$\gamma^2\varsigma^2$ associated with the two gapless modes gives
\begin{equation}\label{app:eq:neutral-gaps}
 \frac{P_{N,\mathrm{boost}}(\varsigma,0,0)}{\gamma^2\varsigma^2}
 =\gamma^2(A_L-C_Lv^2+F_Lv^4)\varsigma^2
 +\gamma(B_L-D_Lv^2)\varsigma+w^2(1-xv^2).
\end{equation}
The leading coefficient is positive because
\begin{equation}\label{app:eq:neutral-leading}
 A_L-C_Lv^2+F_Lv^4
       =A_L(1-y_-v^2)(1-y_+v^2)>0\qquad (|v|<1).
\end{equation}
Equation~\eqref{app:eq:neutral-design-interlacing} gives
$y_-\leq R\leq y_+\leq1$, so the linear coefficient is positive as well:
$B_L-D_Lv^2=B_L(1-Rv^2)>0$. Finally, $0<x<1$ makes the constant term in
\eqref{app:eq:neutral-gaps} positive. All three coefficients
are therefore positive, proving that both gaps have negative real parts,
including at luminal saturation $y_+=1$.

\subsubsection*{Stability at every nonzero wavevector}
We fix real $k_z,k_\perp$ with $k^2=k_z^2+k_\perp^2>0$.
Either component may vanish, so this includes propagation parallel and
perpendicular to the boost. Suppose that a root were on the imaginary
axis of the growth-rate plane, $\varsigma=i\sigma$ with
$\sigma\in\mathbb R$. Equation~\eqref{app:eq:neutral-boost} then gives
\begin{equation}\label{app:eq:boost-imaginary-root}
 \varsigma'=i\Omega,\quad \Omega=\gamma(\sigma+vk_z),\qquad
 K^2=q'=k_\perp^2+\gamma^2(k_z+v\sigma)^2\geq0.
\end{equation}
Thus $\Omega$ and $K$ are real. Substitution into
\eqref{eq:neutral-sound-polynomial} gives
\begin{align}
 \Im P_N(i\Omega,K^2)&=\Omega(-B_L\Omega^2+D_LK^2),\\
 \Re P_N(i\Omega,K^2)&=A_L\Omega^4-(w^2+C_LK^2)\Omega^2
                     +xw^2K^2+F_LK^4.
\end{align}
For a root, both parts must vanish. The imaginary part allows only
$\Omega=0$ or $\Omega^2=RK^2$, since $R=D_L/B_L$.
The possibility $K=\Omega=0$ is excluded: invertibility of the boost
would give $\sigma=k_z=k_\perp=0$, contrary to $k>0$.
If $\Omega=0$, we therefore have $K>0$ and
$\Re P_N=xw^2K^2+F_LK^4>0$. If $\Omega^2=RK^2$, then
\eqref{app:eq:neutral-design-interlacing} instead gives
\begin{equation}\label{app:eq:neutral-axis-exclusion}
 \Re P_N=w^2(x-R)K^2-A_L\Lh K^4<0.
\end{equation}
Here $R>x$ and $\Lh\geq0$ ensure the strict inequality. Neither case
allows a root on the imaginary axis.

It remains to exclude roots entering the right half-plane from infinity.
For the fixed wavevector, we vary $v$ from zero to any chosen
$v_*$ with $|v_*|<1$. Writing the boosted quartic as
$\sum_{j=0}^4a_j(v)\varsigma^j$, its leading coefficient is
\[
 a_4(v)=\gamma^4A_L(1-y_-v^2)(1-y_+v^2)>0.
\]
It has a positive lower bound on the closed interval between $0$ and
$v_*$. The other coefficients are continuous and bounded there.
The polynomial root bound
\begin{equation}\label{app:eq:boost-root-bound}
 |\varsigma|\leq1+\max_{0\leq j<4}|a_j(v)/a_4(v)|
\end{equation}
therefore bounds all roots uniformly along this interval. Since roots
can neither cross the imaginary axis nor escape to infinity, all four
remain in the left half-plane as $v$ varies from $0$ to $v_*$. The
wavevector and final boost were arbitrary. Together with the $k=0$
gap calculation, this proves stability for every real wavevector and
every finite boost.

\subsubsection*{Causality for all propagation angles}
For either squared characteristic speed $y=y_\pm\in[0,1]$, the
corresponding factor of the principal polynomial
\eqref{app:eq:neutral-principal} gives
$\omega'^2=y|\boldsymbol k'|^2$ in the fluid rest frame.
To express this in the moving frame, we use the Lorentz transformations
underlying \eqref{app:eq:neutral-boost},
\[
 \omega'=\gamma(\omega-vk_z),\qquad
 k_z'=\gamma(k_z-v\omega),\qquad k_\perp'=k_\perp.
\]
Writing $k_z=k\cos\phi$, $k_\perp=k\sin\phi$ and $\omega=ck$,
with $k>0$, and dividing the rest-frame equation by $\gamma^2k^2$ gives
\[
 (c-v\cos\phi)^2
 =y\bigl[(\cos\phi-vc)^2+(1-v^2)\sin^2\phi\bigr].
\]
Expanding both sides gives the quadratic equation for the characteristic
velocity $c$ in the moving frame,
\begin{equation}\label{app:eq:neutral-speed-quadratic}
 (1-yv^2)c^2-2v(1-y)\cos\phi\,c
 +v^2\cos^2\phi-y(1-v^2+v^2\cos^2\phi)=0.
\end{equation}
Its roots are
\begin{equation}\label{app:eq:neutral-speed-boosted}
 c_\pm(y,\phi)=\frac{v(1-y)\cos\phi
 \ \pm\sqrt{y(1-v^2)[1-v^2+v^2(1-y)\sin^2\phi]}}{1-yv^2}.
\end{equation}
The discriminant is nonnegative and $1-yv^2>0$. Moreover the characteristic
covector obeys
\[
 |\boldsymbol k|^2-\omega^2
 =|\boldsymbol k'|^2-\omega'^2=(1-y)|\boldsymbol k'|^2\geq0,
\]
so $|c_\pm|\leq1$. At $y=0$ the root is the advective speed $v\cos\phi$.
At $y=1$ the roots are exactly $\pm1$. Applying this calculation to both
$y_\pm$ completes the causality proof in Proposition~\ref{app:prop:neutral}.

\subsubsection{Boosted shear and charge factors}
\label{app:sec:shear}
Both the shear and charge factors have the form
\[
 Q(\varsigma,q)=a\varsigma^2+b\varsigma+dq,
 \qquad a\geq d>0,\quad b>0,
\]
where $a,b,d$ are constant coefficients at the equilibrium state.
At rest, positivity gives two damped roots for $q>0$. After
\eqref{app:eq:neutral-boost}, the leading coefficient is
$\gamma^2(a-dv^2)>0$, and the $k=0$ roots are $0$ and
$-b/[\gamma(a-dv^2)]<0$.
For $k>0$, an imaginary-axis root would satisfy
\[
 \Im Q(i\Omega,K^2)=b\Omega=0,\qquad
 \Re Q(0,K^2)=dK^2>0,
\]
where $\Omega,K$ are defined in \eqref{app:eq:boost-imaginary-root}.
The second inequality follows from $K>0$ when $\Omega=0$ and $k>0$,
and excludes such a root. The leading coefficient is bounded away
from zero on each closed finite boost interval, so the root-bound
argument of Section~\ref{app:sec:neutral-criterion} excludes roots
entering from infinity. Both roots therefore remain damped at every
nonzero real wavevector. Causality follows from
\eqref{app:eq:neutral-speed-boosted} with $y=d/a\leq1$.

For the shear sector, $(a,b,d)=(\theta_1,w,\eta)$, so the required
bound is precisely \eqref{eq:strategy-shear-bound}. Each transverse
polarization has the gap
\begin{equation}\label{app:eq:shear-boosted-gap}
 \varsigma_{T,\mathrm{gap}}=-\frac{w}{\gamma(\theta_1-\eta v^2)}<0.
\end{equation}
No relation between $\theta_1$ and $\theta_2$ is used, and the proof
includes the luminal endpoint $\theta_1=\eta$.\footnote{\label{fn:shear-zero-transport}At
$\eta=0$, the shear polynomial becomes
$\varsigma'(\theta_1\varsigma'+w)$. It has an undamped advected mode
and, if $\theta_1>0$, a damped mode
$\varsigma=-ivk_z-w/(\gamma\theta_1)$.
If $\theta_1=0$, only advection remains. These modes are causal and
linearly stable, but their comoving characteristic speed is zero.}

For the charge sector, we take $\mathcal D>0$ and
$\tau_q=\alpha_qD_q\geq D_q>0$. The same proof applies with
$(a,b,d)=(\tau_q,1,D_q)$, establishing stability at all real
wavevectors and the squared comoving characteristic speed
$y_q=D_q/\tau_q\leq1$.

\subsubsection{Stability and causality of the full boosted charged system}
\label{app:sec:charged-boost-proof}\label{app:sec:charged-numbered}
We now apply the charged uplift to a neutral seed satisfying
Proposition~\ref{app:prop:neutral}, with $\Gamma>0$ and
$\theta_1\geq\eta>0$. We first take $\mathcal D>0$, so
$D_q>0$, and choose $\tau_q=\alpha_qD_q$ with
$\alpha_q\geq1$.\footnote{\label{fn:charged-zero-transport}At
$\mathcal D=0$ with finite $\alpha_q$, $D_q=\tau_q=0$ and
$F_q=\varsigma'$. Charge is then advected without damping,
$\varsigma=-ivk_z$, and the charge relaxation gap is absent.
For vanishing viscosities, the uplift multiplies the reduced neutral
sound polynomial of footnote~\ref{fn:neutral-zero-transport}, evaluated
at the charged values $\zeta_*$, by $-(\Delta_J/w)F_q$.
In particular, $\Gamma=0$ implies
$\eta=\zeta=0$ and $\mathcal D B^2=0$. The remaining modes are ideal sound, an additional damped mode for \(F_3\) when \(\varepsilon_1>0\), and the charge
telegrapher modes if $\mathcal D>0$, or charge advection if
$\mathcal D=0$. Their reduced polynomials establish linear stability
and causality in every inertial frame.}
The factorization \eqref{eq:factorization} holds for complex
$\varsigma,q$ and therefore remains valid after the simultaneous
substitution \eqref{app:eq:neutral-boost}. The sound factor is covered
by Section~\ref{app:sec:neutral-criterion}, and both the shear and charge
factors by Section~\ref{app:sec:shear}. Together these results establish
stability at nonzero real wavevectors. We now extract the combined
$k=0$ gaps and collect the causality results for the full system.

At $k=0$, we extract the gaps directly from the boosted determinant.
For the charge factor, \eqref{app:eq:neutral-boost} gives
$F_q(\gamma\varsigma,-\gamma^2v^2\varsigma^2)
=\gamma\varsigma(1+\ell_q\varsigma)$.
The sound factor is $\gamma^2\varsigma^2$ times the quadratic
in \eqref{app:eq:neutral-gaps}. The coefficients in these two factors are
\begin{align}\label{app:eq:charged-gap-coefficients}
 \ell_q&=\gamma(\tau_q-D_qv^2)>0,&
 a_g&=\gamma^2(A_L-C_Lv^2+F_Lv^4)>0,\nonumber\\
 b_g&=\gamma(B_L-D_Lv^2)>0,&c_g&=w^2(1-xv^2)>0.
\end{align}
Their positivity follows from $\tau_q\geq D_q>0$ and the sound-gap
calculation above. Multiplying the two factors gives
\begin{equation}\label{app:eq:charged-uplift-gaps}
 \frac{P_{\rm ch,boost}(\varsigma,0)}{\gamma^3\varsigma^3}
 =-\frac{\Delta_J}{w}(1+\ell_q\varsigma)(a_g\varsigma^2+b_g\varsigma+c_g).
\end{equation}
The linear factor gives the negative gap $-1/\ell_q$.
The quadratic has positive coefficients $a_g,b_g,c_g$, so both of
its roots have negative real parts. This proves stability of all
three longitudinal gaps directly from the original boosted determinant.
The extracted factor $\varsigma^3$ accounts for the three zero-frequency
longitudinal modes. The two shear gaps are given by
\eqref{app:eq:shear-boosted-gap}.

For causality, the squared speeds in the fluid rest frame are
$y_-,y_+,y_T,y_q$, as listed in Table~\ref{tab:families}.
Applying \eqref{app:eq:neutral-speed-boosted} to each gives real
characteristic velocities with $|c|\leq1$ in every finite boosted frame.
When $y<1$ the inequality is strict, including the advective branch
$c=v\cos\phi$ at $y=0$. At $y=1$ the velocities are $\pm1$.\footnote{\label{fn:deformation-boosts}The shifts \eqref{eq:free-shifts} preserve these results. Substituting the boosted arguments \eqref{app:eq:neutral-boost} into \eqref{app:eq:hierarchy-isospectral} gives identical boosted dispersion polynomials at every real wavevector and finite boost. At $k=0$, $\varsigma'=\gamma\varsigma$ and $q'=-\gamma^2v^2\varsigma^2$, so removing the common factor $\gamma^3\varsigma^3$ also gives identical gap polynomials. The frequency roots, gaps and characteristic speeds are therefore unchanged.}

\section{Comments on existing frames}\label{app:existing-frames}
The constitutive relations of Ref.~\cite{BDN2022} provide another route to
stable and causal frames with prescribed physical transport. In this appendix, we show this
directly for neutral fluids and for charged fluids at nonzero density, and comment on
the obstructions for zero density states.

\subsection{Constitutive class and neutral construction}
\label{app:existing-neutral}
In our notation, the charged class of Ref.~\cite{BDN2022} takes the form
\begin{align}
 \mathcal E&=\tau_\epsilon(D\epsilon+w\Theta),&
 \mathcal P&=\tau_P(D\epsilon+w\Theta)-\zeta\Theta,\nonumber\\
 \mathcal Q^\mu&=\tau_Q(w a^\mu+\perpgrad^\mu p)
       +\frac{\cD T w}{n}\perpgrad^\mu\alpha,&
 J^\mu&=nu^\mu.
 \label{app:eq:existing-class}
\end{align}
The three relaxation parameters $\tau_\epsilon,\tau_P,\tau_Q$ parametrise the frame choice. The coefficients $\eta$, $\zeta$ and
$\cD$ remain physical inputs. Using $T\epsilon_T=w/x$,
the neutral coefficients become
\begin{align}
 \varepsilon_1&=\frac{w\tau_\epsilon}{x},&
 \varepsilon_2&=w\tau_\epsilon,&
 \pi_1&=\frac{w\tau_P}{x},\nonumber\\
 \pi_2&=w\tau_P-\zeta,&
 \theta_1&=\theta_2=w\tau_Q.
 \label{app:eq:existing-neutral-map}
\end{align}
For both the
neutral and charged calculations below, we write $d=(4\eta/3+\zeta)/w$,
using the physical bulk viscosity. In the neutral case this is
$\Gamma/w$, with $\Gamma$ as in \eqref{eq:quartic-data}, and
a sufficient prescription in order to attain causal and stable fluids for generic EoS and prescribed transport is
\begin{equation}\label{app:eq:existing-neutral-choice}
 \tau_\epsilon=\tau_Q=\tau,\qquad \tau_P=x\tau,\qquad
 \tau=d\left[\frac1x+\frac{1+x}{(1-x)^2}\right].
\end{equation}
For $d>0$, all relaxation parameters are positive. Substitution into the
original matrix \eqref{app:eq:neutral-matrix} gives the normalized
polynomial $\widehat P_N=P_N/w^2=x\det M/w^2$:
\begin{align}
 \widehat P_N(\varsigma,q)
 ={}&\tau^2\varsigma^4+2\tau\varsigma^3
 +[1+(2x\tau^2+d\tau)q]\varsigma^2\nonumber\\
 &+(2x\tau+d)q\varsigma+xq+(x^2\tau^2-xd\tau)q^2.
 \label{app:eq:existing-neutral-polynomial}
\end{align}
Since $\tau>d/x$, its coefficients are positive for $q>0$.
The Hurwitz determinants of this normalized polynomial are
\begin{align}
 \Delta_1&=2\tau,\qquad
 \Delta_2=2\tau+\tau^2(2x\tau+d)q,\nonumber\\
 \Delta_3&=\tau d q[2+\tau(8x\tau+d)q],\nonumber\\
 \Delta_4&=[xq+(x^2\tau^2-xd\tau)q^2]\Delta_3.
 \label{app:eq:existing-neutral-hurwitz}
\end{align}
They are strictly positive for every $q>0$. At $q=0$,
$\widehat P_N=\varsigma^2(\tau\varsigma+1)^2$. The squared longitudinal characteristic speeds are
\begin{equation}\label{app:eq:existing-neutral-speeds}
 y_\pm=x+\frac{\delta_{\tau}}2\pm\frac12\sqrt{\delta_{\tau}^2+8x\delta_{\tau}},
 \qquad \delta_{\tau}=\frac d\tau.
\end{equation}
The prescription gives $0<\delta_{\tau}<x$ and
$\delta_{\tau}<(1-x)^2/(1+x)$. The speed polynomial has positive product
$x(x-\delta_{\tau})$, positive sum $2x+\delta_{\tau}<2$, and positive value
$(1-x)^2-\delta_{\tau}(1+x)$ at squared speed one. Its distinct real roots
therefore obey $0<y_-<y_+<1$. The shear polynomial, divided by $w$,
is $\tau\varsigma^2+\varsigma+(\eta/w)q$. Its squared speed
$\eta/(w\tau)<1$ follows from $\tau>d/x\geq4\eta/(3wx)$ when
$\eta>0$. At $\eta=0$ it factors as
$\varsigma(\tau\varsigma+1)$, retaining an advective root.
We discuss the case of arbitrary boosts in Section~\ref{app:existing-boosts}.

If $d=0$, both viscosities vanish and \eqref{app:eq:existing-neutral-choice}
sets all frame coefficients to zero. The reduced system has ideal sound
$\varsigma^2+xq=0$ and advected transverse perturbations. 
At a regular conformal
limit $x\to1/3$, $\zeta\to0$, the prescription instead gives the finite
values $\tau\to8\eta/w$, $\varepsilon_1\to24\eta$,
$\varepsilon_2,\pi_1\to8\eta$, $\pi_2\to8\eta/3$ and
$\theta_1,\theta_2\to8\eta$, which obey \eqref{eq:conformal}.

The class of frames introduced in
\eqref{app:eq:existing-neutral-map} requires
$\varepsilon_2=x\varepsilon_1$. Entropy matching \eqref{eq:matching} then gives
$\pi_2=x\pi_1-\zeta$. The subset of \family{4} specified in
footnote~\ref{fn:F4-existing-subset} satisfies these relations. On the
 domain $\eta>0$, \family{1}-\family{3} do not, while
\family{5} and \family{6} satisfy them at $x=1/3$ only. 

\subsection{Charged construction at nonzero density}
\label{app:existing-charged}
In the charged case and for $n\ne0$, the coefficients of \eqref{app:eq:existing-class} in
our basis are
\begin{align}
 (\varepsilon_1,\varepsilon_2,\varepsilon_3)
   &=\tau_\epsilon(TE,w,F),\nonumber\\
 (\pi_1,\pi_2,\pi_3)&=(\tau_PTE,\tau_Pw-\zeta,\tau_PF),\nonumber\\
 \theta_1=\theta_2&=\tau_Qw,\qquad
 \theta_3=\tau_QrwT+\frac{\cD T}{r},\qquad
 \nu_i=\gamma_i=0.
 \label{app:eq:existing-charged-map}
\end{align}
Equations~\eqref{eq:matching} give precisely the prescribed $\zeta$ and
$\cD$. All identities in \eqref{eq:thermo-domain} hold. With $d$ as above, we introduce
\begin{align}
 g_I&=\frac{\cD TB}{rw},\qquad
 h_I=\frac{\cD TE}{\Delta_J},\nonumber\\
 K_I&=x(h_I-D_q)=(x-A)g_I=\frac{\cD T^2B^2}{w}\geq0.
 \label{app:eq:existing-auxiliary}
\end{align}
Here $xD_q=\cD w/\Delta_J$ follows from \eqref{eq:effective}.
The remaining identities use $E=w/(Tx)+T\Delta_JB^2/(wx)$ and
\eqref{eq:thermo}. In particular $h_I,D_q\geq0$, whereas $A$ and
$g_I$ can have either sign. For charged fluids, the $\Gamma$ used in
our uplift satisfies $\Gamma/w=d+K_I/x$, whereas $d$ contains only
the physical viscosities.
For the original matrix \eqref{app:original-charged-matrix}, direct
expansion gives $\det M_{\rm charged}=-\Delta_Jw P_I$, with
\begin{align}
 P_I(\varsigma,q)={}&\tau_\epsilon\tau_Q\varsigma^5
 +(\tau_\epsilon+\tau_Q)\varsigma^4
 +(1+\mathsf Hq)\varsigma^3+\mathsf Dq\varsigma^2\nonumber\\
 &+(xq+\mathsf Cq^2)\varsigma+xD_qq^2,\label{app:eq:existing-quintic}\\
 \mathsf H&=\tau_\epsilon(x\tau_Q+d+g_I)+\tau_P\tau_Q,\nonumber\\
 \mathsf D&=x(\tau_\epsilon+\tau_Q)+d+h_I,\nonumber\\
 \mathsf C&=\tau_P(x\tau_Q+g_I)-d(\tau_QA+g_I-h_I).
 \label{app:eq:existing-quintic-coefficients}
\end{align}
The shear polynomial is $\tau_Q\varsigma^2+\varsigma+(\eta/w)q$. We can pick any positive time scale $\ell$ and set
\begin{align}
 a_{\star}&=1+(A/x)^2,\qquad
 y_{\star}=x-x(1-x)\frac{g_I\ell}{\ell^2+g_I^2},\nonumber\\
 \tau_Q&=\tau_\star,\qquad
 \tau_\epsilon=a_{\star}\tau_\star,\quad
 \tau_P=a_{\star}y_{\star}\tau_\star.
 \label{app:eq:existing-charged-choice}
\end{align}
Our aim is to choose a finite, sufficiently large $\tau_\star>0$, without
changing the physical inputs. Since
$|g_I\ell/(\ell^2+g_I^2)|\leq1/2$, we have $0<y_{\star}<1$ and
\begin{align}
 g_I(x-y_{\star})&=\frac{x(1-x)\ell g_I^2}{\ell^2+g_I^2}\geq0,\nonumber\\
 S_\star&=a_{\star}x+A>0,\qquad M_\star=y_{\star}+a_{\star}x+A(1+1/a_{\star})>0.
 \label{app:eq:existing-positive-quantities}
\end{align}
Indeed, $S_\star=x[(A/x+1/2)^2+3/4]$. For $A\geq0$, positivity
of $M_\star$ is immediate. For $A<0$, write $z_A=-A/x>0$. Then
$a_{\star}=1+z_A^2\geq2z_A$ gives
$M_\star\geq y_{\star}+x(a_{\star}-1)/2>0$. Consequently
\begin{equation}\label{app:eq:existing-Z}
 Z_\star=dxM_\star+g_I(x-y_{\star})S_\star>0,\qquad dx+K_I>0,
 \quad\hbox{if }d>0\hbox{ or }g_I\ne0.
\end{equation}

First assume $d>0$ or $g_I\ne0$. For each of the first four Hurwitz
determinants of \eqref{app:eq:existing-quintic}, the table lists every
nonzero coefficient of a power of $q$, its degree in $\tau_\star$, and
its leading coefficient in which the relation $xh_I=xD_q+(x-A)g_I$ has been used.
\begin{center}
\begin{tabular}{cccc}
\toprule
Determinant & Power of $q$ & Degree in $\tau_\star$ & Leading coefficient\\
\midrule
$\Delta_1$ & $0$ & $1$ & $a_{\star}+1$\\
$\Delta_2$ & $0$ & $1$ & $a_{\star}+1$\\
$\Delta_2$ & $1$ & $3$ & $a_{\star}y_{\star}(a_{\star}+1)$\\
$\Delta_3$ & $1$ & $1$ & $(a_{\star}+1)(d+h_I)$\\
$\Delta_3$ & $2$ & $3$ & $a_{\star}(a_{\star}+1)(Z_\star+xy_{\star}D_q)/x$\\
$\Delta_4$ & $2$ & $1$ & $(a_{\star}+1)(dx+K_I)$\\
$\Delta_4$ & $3$ & $3$ & $a_{\star}(a_{\star}+1)[Z_\star+y_{\star}(dx+K_I)]$\\
$\Delta_4$ & $4$ & $5$ & $a_{\star}^2y_{\star}(a_{\star}+1)Z_\star$\\
\bottomrule
\end{tabular}
\end{center}
All entries in the last column are positive. There are only finitely
many coefficient polynomials, so one finite $\tau_\star$ makes all of
them positive simultaneously, uniformly for every $q>0$ at the given
state. If $\cD>0$, then $D_q>0$ and
$\Delta_5=xD_qq^2\Delta_4>0$, proving strict rest-frame stability.
If $\cD=0$, then $g_I=h_I=D_q=K_I=0$ and
$P_I=\varsigma P_4$. The first four Hurwitz determinants are those of
$P_4$ and remain positive for $d>0$, while the exact advective root is
retained. At $q=0$,
$P_I=\varsigma^3(\tau_\epsilon\varsigma+1)(\tau_Q\varsigma+1)$.

To choose a sufficiently large $\tau_\star$, we first fix the physical inputs
and choose any reference time $\ell>0$. We write $\tau_\star=\ell z$, so that only
the dimensionless parameter $z$ remains to be chosen. Each required
inequality takes the form $p_\star(z)>0$, where
$p_\star(z)=c_mz^m+\cdots+c_0$ has a positive leading coefficient $c_m$.
For large $z$, the positive highest-power term dominates the remaining
terms. An explicit sufficient bound is therefore
\[
 z>1+\frac{1}{c_m}\sum_{j=0}^{m-1}|c_j|.
\]
\begin{samepage}
For example, labelling these finitely many polynomials by $i$ and writing
$p_i(z)=\sum_{j=0}^{m_i}c_{ij}z^j$, where $m_i$ is the degree at the
chosen state and $c_{i m_i}>0$, an explicit sufficient choice is
\[
\tau_\star=\ell\left[
 2+\max_i\sum_{j=0}^{m_i-1}\frac{|c_{ij}|}{c_{i m_i}}
 \right].
\]
The maximum includes the coefficient inequalities in the table and the
characteristic and gap inequalities below. This choice makes
$z=\tau_\star/\ell$ strictly larger than every required bound, so all
these conditions hold for the prescribed inputs. The reference time
$\ell$ remains fixed. The resulting $\tau_\star$ is finite at the chosen
state, although it need not be the smallest admissible value.
\end{samepage} 
Writing $y=c^2$ as in \eqref{eq:speeds}, the principal sound speeds
squared are the roots of
\begin{align}
 &a_{\star}\tau_\star^2 y^2-\mathsf Hy+\mathsf C=0,\nonumber\\
 &\mathsf H=a_{\star}(x+y_{\star})\tau_\star^2+a_{\star}(d+g_I)\tau_\star,\qquad
 \mathsf C=a_{\star}xy_{\star}\tau_\star^2+(a_{\star}y_{\star}g_I-Ad)\tau_\star-d(g_I-h_I).
 \label{app:eq:existing-charged-speed-polynomial}
\end{align}
Their discriminant is
\begin{align}
 \mathsf H^2-4a_{\star}\tau_\star^2\mathsf C={}&
 a_{\star}^2(x-y_{\star})^2\tau_\star^4
 +2a_{\star}\{d[a_{\star}(x+y_{\star})+2A]+a_{\star}g_I(x-y_{\star})\}\tau_\star^3\nonumber\\
 &+\{a_{\star}^2(d+g_I)^2+4a_{\star}d(g_I-h_I)\}\tau_\star^2.
 \label{app:eq:existing-charged-discriminant}
\end{align}
The cubic coefficient is positive: $a_{\star}(x+y_{\star})+2A>0$ follows, for negative
$A$, from $a_{\star}\geq2|A|/x$, and $g_I(x-y_{\star})>0$ if $g_I\ne0$.
The discriminant is therefore positive for sufficiently large
$\tau_\star$, also when $x=y_{\star}$ and its quartic coefficient vanishes.
For large $\tau_\star$, the inequalities
\begin{equation}\label{app:eq:existing-speed-bounds}
 \mathsf C>0,\qquad \mathsf H>0,\qquad
 2a_{\star}\tau_\star^2-\mathsf H>0,\qquad
 a_{\star}\tau_\star^2-\mathsf H+\mathsf C>0
\end{equation}
follow from their positive leading coefficients. The last is
$a_{\star}(1-x)(1-y_{\star})$. As in the neutral case, these give two distinct squared
speeds $0<y_-<y_+<1$, together with the advective characteristic.
Taking $\tau_\star>\eta/w$ also makes the shear cone subluminal.

\subsection*{Cases with vanishing transport}

The case $d=g_I=0$ is not covered by the strict Hurwitz argument above.
It implies $\eta=\zeta=0$ and $D_q=h_I$. Direct factorisation of the
original quintic gives
\begin{equation}\label{app:eq:existing-boundary-factorisation}
 P_I=(\varsigma^2+xq)
 \left[a_{\star}\tau_\star^2\varsigma^3+(a_{\star}+1)\tau_\star\varsigma^2
 +(1+a_{\star}y_{\star}\tau_\star^2q)\varsigma+h_Iq\right].
\end{equation}
Here the prescription has $y_{\star}=x$. The first factor is an undamped
causal sound wave. If $h_I>0$, the cubic has positive coefficients for
$q>0$ and its remaining Hurwitz condition is
\begin{equation}\label{app:eq:existing-boundary-margin}
 (a_{\star}+1)\tau_\star+
 a_{\star}\tau_\star^2[(a_{\star}+1)y_{\star}\tau_\star-h_I]q>0,
\end{equation}
which holds whenever $\tau_\star>h_I/[(a_{\star}+1)y_{\star}]$. If $h_I=0$, the cubic
instead factors as
$\varsigma[a_{\star}\tau_\star^2\varsigma^2+(a_{\star}+1)\tau_\star\varsigma
+1+a_{\star}y_{\star}\tau_\star^2q]$. The quadratic is strictly stable and the
advective root remains. The principal squared speeds are $x$, $y_{\star}$
and zero, with coincident sound speeds allowed in this case.
At $\eta=0$, each shear factor is
$\varsigma(\tau_\star\varsigma+1)$. These exact reduced factors
establish spectral stability and causal characteristics.

\subsection{Generalisations to arbitrary boosts}
\label{app:existing-boosts}
We use the Lorentz substitution \eqref{app:eq:neutral-boost}, keeping
the wavevector real in the moving frame to check stability and causality at arbitrary boost. For the neutral polynomial,
direct evaluation at $k=0$ gives
\begin{align}
 \frac{\widehat P_{N,\mathrm{boost}}}{\gamma^2\varsigma^2}
 ={}&1-xv^2+\gamma[2\tau-(2x\tau+d)v^2]\varsigma\nonumber\\
 &+\gamma^2[\tau^2-(2x\tau^2+d\tau)v^2
 +(x^2\tau^2-xd\tau)v^4]\varsigma^2.
 \label{app:eq:existing-neutral-boost-gap}
\end{align}
The quadratic coefficient is
$\gamma^2\tau^2(1-y_+v^2)(1-y_-v^2)>0$. The linear
bracket is bounded below by
\begin{equation}
 2\tau(1-x)-d
 =d\left[\frac{2(1-x)}x+\frac{1+3x}{1-x}\right]>0
 \qquad(d>0).
\end{equation}
Thus both nonzero gaps are damped for every $|v|<1$.
For the charged quintic the corresponding direct calculation is
\begin{align}
 \frac{P_{I,\mathrm{boost}}}{\gamma^3\varsigma^3}
 ={}&1-xv^2+
 \gamma[(a_{\star}+1)\tau_\star-\mathsf Dv^2+xD_qv^4]\varsigma\nonumber\\
 &+\gamma^2[a_{\star}\tau_\star^2-\mathsf Hv^2+\mathsf Cv^4]\varsigma^2.
 \label{app:eq:existing-charged-boost-gap}
\end{align}
The constant and quadratic coefficients are positive, the latter because
its bracket is $a_{\star}\tau_\star^2(1-y_+v^2)(1-y_-v^2)$.
The linear bracket is at least
$(a_{\star}+1)\tau_\star(1-x)-(d+h_I)>0$ after increasing $\tau_\star$ if
necessary. The shear factor at $k=0$ similarly has a negative nonzero
gap, since $\tau_Q-\eta v^2/w>0$.

For a fixed nonzero real wavevector, vary the boost continuously from
zero to any prescribed $|v|<1$. The leading coefficients of the boosted
neutral quartic and charged quintic are, respectively,
\begin{equation}\label{app:eq:existing-boost-leading}
 \gamma^4\tau^2(1-y_+v^2)(1-y_-v^2),\qquad
 \gamma^5a_{\star}\tau_\star^2(1-y_+v^2)(1-y_-v^2).
\end{equation}
Here $y_\pm$ denote the squared speeds of the corresponding neutral or
charged polynomial. These coefficients are nonzero on the compact boost
path, so no root enters from
infinity. A crossing of the imaginary $\varsigma$ axis would give a
real frequency--momentum covector, whose Lorentz transform is also
real. Strict rest-frame stability excludes such a root at nonzero
rest-frame momentum. At zero rest-frame momentum the only imaginary
root is zero, hence the entire covector is zero, which cannot correspond
to a nonzero wavevector in the moving frame. This excludes every
crossing, as in Section~\ref{app:sec:neutral-criterion}.

Marginal factors are separated before this argument is applied.
At $\cD=0$, remove the exact advective factor of $P_I$. The
remaining quartic has nonzero boosted leading coefficient
$\gamma^4a_{\star}\tau_\star^2(1-y_+v^2)(1-y_-v^2)$.
At $d=g_I=0$, remove the causal wave factor in
\eqref{app:eq:existing-boundary-factorisation}. The strictly stable
cubic has boosted leading coefficient
$\gamma^3a_{\star}\tau_\star^2(1-y_{\star}v^2)>0$. If $h_I=0$, also remove its
advective factor, leaving a strictly stable quadratic with leading
coefficient $\gamma^2a_{\star}\tau_\star^2(1-y_{\star}v^2)>0$.
The advective roots and the causal wave roots remain on the imaginary
axis under boosts. The shear factors obey the same argument when
$\eta>0$, with leading coefficient
$\gamma^2(\tau_Q-\eta v^2/w)>0$. When $\eta=0$, their advective and
decaying factors are explicit. Together with the zero-momentum gaps,
this proves stability for all real wavevectors and finite boosts.
All characteristic cones used here are causal, so their boosted
characteristic velocities are also real and bounded by one.

\subsection{Zero density case}
\label{app:existing-zero-density}
The charged construction above requires $n\ne0$. The associated
restriction to baryon-rich matter was already noted in
Ref.~\cite{BDN2022}. The following identity makes its conductivity
obstruction at zero density explicit, without choosing relaxation parameters. Since
$\gamma_i=0$ for the ideal current, the physical conductivity identity
\eqref{eq:matching} reads
\begin{equation}\label{app:eq:existing-zero-density-obstruction}
 \cD T=r(\theta_3-rT\theta_1).
\end{equation}
At $r=0$, finite $\theta_1,\theta_3$ therefore force $\cD=0$.
The $1/n$ term in \eqref{app:eq:existing-class} expresses the same
obstruction. A divergent $\tau_Q$ cannot remove it while keeping
$\theta_1=\tau_Qw$ finite. Thus this class of frames does not cover
the entire charged thermodynamic domain consider here, which includes zero density
with prescribed positive conductivity.

For every fixed regular state with $n\ne0$, the preceding construction
accommodates arbitrary finite prescribed $\eta,\zeta,\cD\geq0$,
without imposing $A>0$ or a relation among the physical transports.
On compact regular state patches with $|n|$ and $d$ bounded away from
zero, the positive Hurwitz leading coefficients admit uniform lower
bounds. The positive cubic coefficient of
\eqref{app:eq:existing-charged-discriminant} controls the discriminant
uniformly even where its quartic coefficient vanishes. All other
coefficients are bounded, so one sufficiently large constant
$\tau_\star$ works throughout such a patch. Smooth constitutive data
then give smooth frame coefficients. Positive shear viscosity suffices
for the stated condition on $d$, while bulk viscosity may vanish.

At a regular charged conformal limit with $n\ne0$ and $d>0$,
$B\to0$, $A,x\to1/3$ give $a_{\star}\to2$, $y_{\star}\to1/3$ and
$\tau_P\to\tau_\epsilon/3$. The relaxation parameters can remain
finite on such a compact patch.  Our
constructions discussed in Section~\ref{sec:overview} supply additional explicit frame choices, including
non-thermodynamic members and regular charged prescriptions at $n=0$.

\phantomsection

\end{document}